# Across Browsers SVG Implementation

**Master's Project**

**Master Computer Science**

**Leiden Institute of Advanced Computer Science**

**Leiden University**

**Liang Wang (0728780)**

**Advisor: Dr. Nies Huijsmans (LIACS)**
**Advisor: Dr. Michael S. Lew (LIACS)**
**Advisor: Dan Tsymbala (Backbase)**

**April. 2009**

# Abstract

When surfing the Internet, a user might observe that the popular image formats displayed are raster images such as jpeg, gif. In addition to those formats, a new technology defined by the W3C called SVG (Scalable Vector Graphics) is bringing rich, compelling, interactive, high-resolution graphics to the Web. SVG offers benefits such as faster download time, scalable images, searchable texts and being compatible with other XML languages. The feature of scaling allows SVG to be viewed in different screen sizes, which makes it more and more widely used in PC displays and handheld devices. Although SVG may be considered a mature technology, most of SVG developers still suffer from its browser compatibility issue – a plug-in must be installed to view SVG images in Internet Explorer. In the meanwhile, it is unlikely that it will be natively supported by Microsoft anytime soon; this may be because Microsoft has its own web vector graphic format called VML. In this Master's project, SVG will be translated into VML or HTML by using Javascript based on Backbase Client Framework. The target of this project is to implement SVG to be viewed in Internet Explorer without any plug-in and work together with other Backbase Client Framework languages. The result of this project will be added as an extension to the current Backbase Client Framework.

# 1. Introduction

SVG (Scalable Vector Graphics) is an XML markup language for describing two-dimensional vector graphics. It is an XML-based language as opposed to a closed binary format (such as Flash). It is explicitly designed to work with other W3C standards such as CSS, DOM and SMIL.

SVG brings the advantages of XML to the world of vector graphics. It enables the textual content of graphics - from logos to diagrams - to be searched, indexed, and displayed in multiple languages. This is a significant benefit for both accessibility and internationalization. However, SVG can not be viewed in Internet Explorer without a plug-in.

VML is another XML language that can draw vector graphics in Internet Explorer. The aim of this project is to use Javascript to translate SVG images to VML images to make them displayable within Internet Explorer without the required plug-in.

There have been efforts to convert SVG images to VML. However, most of the previous implementations are not applicable to drawing complex SVG images as they only converted several simple SVG elements and many important attributes and commands are missing. This project implemented most of the fundamental elements and as a result some very nice images and diagrams.

The final implementation is based on Backbase Client Framework. The result of this implementation is integrated as a binding to the framework, which makes SVG work together with the existing languages in the framework.

The following report is organized according to the following topics:

- WHY SVG?
- WHY USE JAVASCRIPT?
- HOW TO IMPLEMENT INLINE SVG IN NATIVE SUPPORT WEB BROWSERS?
- HOW TO IMPLEMENT SVG IN INTERNET EXPLORER?
- IMPLEMENTATION OF SVG ELEMENTS
- ABSTRACT CLASSES - IMPLEMENTATION OF SVG COMMON ATTRIBUTES
- IMPLEMENTATION OF "TRANSFORM" ATTRIBUTE
- PERFORMANCE TEST
- FUTURE WORK
- CONCLUSION

## 2. Why SVG ?

Why does Backbase Client Framework wants to implement the SVG module? Here are some reasons:

**Great Image Format**
SVG is a vector graphics format. Vector graphics can be scaled without loss of image quality, while raster (bitmap) graphics cannot. Here is an example:

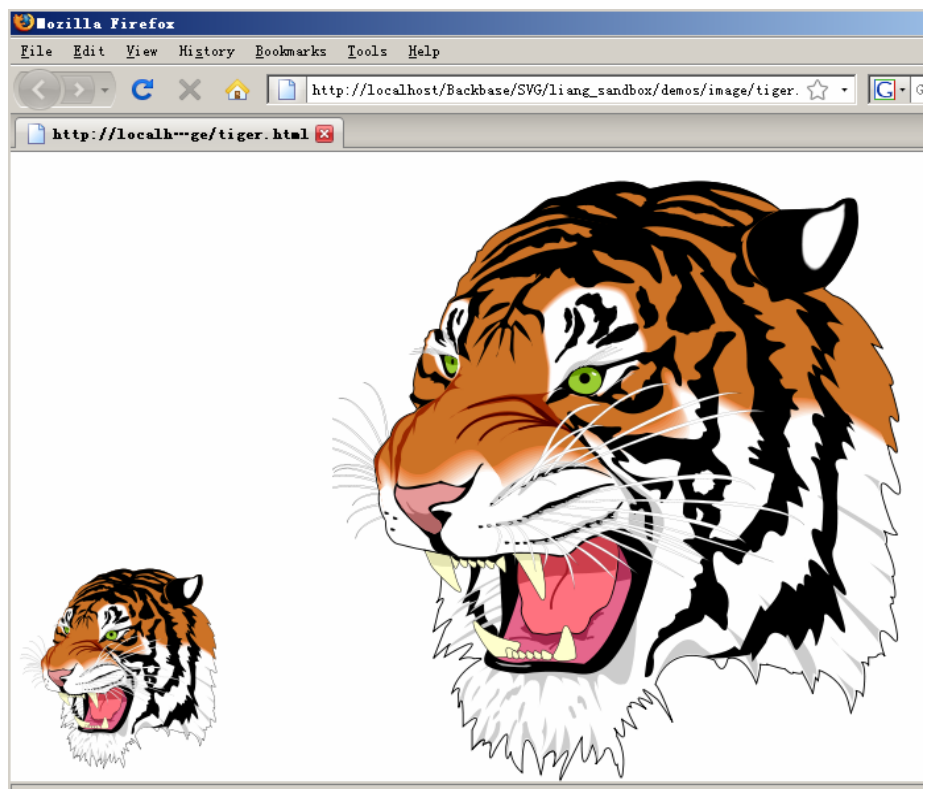

Figure 2-1. SVG Image

**Enhance HTML Content**
SVG works together with HTML, CSS, JavaScript, SMIL and the current Backbase Client Framework languages such as XEL. They use SVG to enhance a regular HTML page or web application.

**Animation and Interactions**
SVG has a document model (DOM) which is accessible from JavaScript and XEL. This allows developers to create rich animations and interactive images.

**Mapping, Charting, Games & 3D Experiments**
A little SVG can go a long way to enhanced web content. Developers can use SVG to create nice diagrams which normal HTML can not do. Here is another

example:

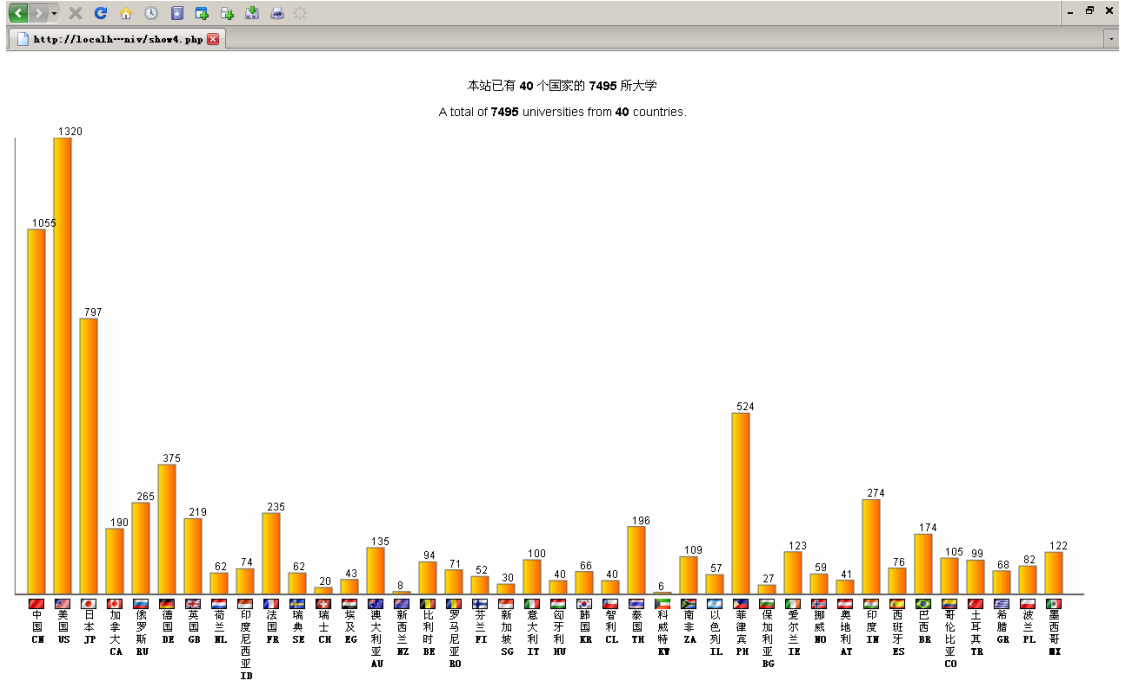

**Figure 2-2. SVG Diagram**

# 3. Why use Javascript ?

There are several approaches to implement SVG in Backbase Client Framework. The possible approaches are: Javascript, XSLT (Extensible Stylesheet Language Transformations), HTC (HTML Components).

One of most important aspects we should take into account is the performance of these approaches. I have done some testing work on all these three approaches and Javascript proved to be the fastest one.

Another factor is that the size of the implementation file should be as small as possible. I found two existing libraries (*VectorConverter and svg-in-ie*), both of which have implemented a small sub-set of SVG. The following comparison is based on the two libraries.

| Library | Type | *Coverage | Current File Size | *Estimated File Size |
|---------|------|-----------|-------------------|----------------------|
| VectorConverter | XSLT | 80% | 582 KB | 726KB |
| svg-in-ie | Javascript | 5% | 18 KB | 360KB |

(*Coverage – How much SVG elements have been implemented in the library

*Estimated File Size – Estimate how large the file would be if all the SVG tags are implemented.

This file size is calculated based on the current file size and the coverage.

That is *Estimated File Size = Current File size/Coverage*)

**Table 3-1. Comparison of libraries**

From the table, we can see that Javascript has an obvious advantage to XSLT when considering the file size.

This paper is not about how to select an appropriate approach to implement SVG, so I will not explain it in more detail. For more information about how to select the ideal approach, please refer to my research project paper "*How To Select an Implementation for Vector Graphics Across Browsers*".

# 4. How to implement inline SVG in native support web browsers ?

Standard web browsers such as Firefox and Safari support SVG natively, which means you don't have to install any third-party plug-ins into the browser when you view the SVG images. In this chapter, all the problems are discussed based on standard browsers.

## 4.1 SVG in XHTML or XML files

If users wish to display SVG images in XHTML or XML files, they can write SVG code directly in the file which we called *inline* SVG. A simple example is as follow:

```
<html xmlns="http://www.w3.org/1999/xhtml">
  <head>
    <meta http-equiv="Content-Type" content="text/html; charset=utf-8"/>
    <title>Example of SVG</title>
  </head>
  <body>
      <svg:svg xmlns:svg=http://www.w3.org/2000/svg
               viewBox="0 0 800 800" width="800" height="800">
          <svg:rect x="1" y="1" width="1198" height="398"
                    fill="red" stroke="blue" stroke-width="2" />
      </svg:svg>
  </body>
</html>
```
**Example 4-1. Source code of demo.xhtml and demo.xml**

## 4.2 SVG in HTML files

When users want to present SVG images in HTML files, unlike in XML or XHTML, they must use `<object>` or `<embed>` element to embed the SVG files into the HTML files (I will discuss the reason later) which we called *referenced* SVG, a simple example can be as follow:

```
<html xmlns="http://www.w3.org/1999/xhtml">
    <head>
        <meta http-equiv="Content-Type" content="text/html; charset=utf-8"/>
        <title>Example of SVG</title>
    </head>
    <body>
        <object data="demo.svg" type="image/svg+xml"
                width="640px" height="480px"/>
    </body>
</html>
```

**Example 4-2. Source code of demo.html**

```
<svg:svg viewBox="0 0 800 800" width="800" height="800"
         xmlns:svg="http://www.w3.org/2000/svg">
    <svg:rect x="1" y="1" width="1198" height="398"
              fill="red" stroke="blue" stroke-width="2" />
</svg:svg>
```

**Figure 4-3. Source code of demo.svg**

## 4.3 Comparison between inline and referenced SVG

Most of the web developers would prefer inline SVG to referenced SVG. The reason is very simple - inline SVG is more interactive. Developers could manipulate DOM very easily with inline SVG. Moreover, it's very easy to have inline SVG interact with server side. For instance, developers can use PHP, JSP or other server side languages to create SVG elements dynamically; it's also possible for developers to use SVG in AJAX applications.

On the other hand, referenced SVG is not flexible enough. As in referenced SVG, the SVG code is separated from the HTML code; it's hardly possible for developers to interact with a SVG file within a HTML file. So the developers can never manipulate or create SVG dynamically but write the SVG file in advance.

Therefore, it's a high pressure to make inline SVG available in all kinds of web page files.

## 4.4 Why does not inline SVG work in HTML?

You may have already found out that if you change the file extension of the previous demo files demo.xml or demo.xhtml to demo.html, nothing will display when you open it in standard browsers even the source code is identical.

The reason for this is that SVG is a language in XML. SVG (XML) code can be parsed and executed by XML parsers of web browsers. When the browsers load web page files, it will judge the contents of the files by their extensions. Specifically, when loading a file namely demo.html, the browser will consider the content of this file to be HTML code by default, similarly, if a file demo.xml is loaded, the content of it will be considered as XML code.

That is to say, even if you write some SVG (XML) code in a HTML file, the browser will still consider these SVG (XML) code to be HTML. The example below explains this in details.

Supposing that we have a file with the code in Example 4-4,

```
<html>
<head></head>
  <body>
    <svg:svg viewBox="0 0 800 800" width="800" height="800"
            xmlns:svg="http://www.w3.org/2000/svg">
        <svg:rect x="1" y="1" width="1198" height="398"
                fill="red" stroke="blue" stroke-width="2" />
    </svg:svg>
  </body>
</html>
```

**Example 4-4. demo code**

If the file is save as a XML or XHTML file, when the browser load this file, elements `<svg:svg>` and `<svg:rect>` will be recognized as `<svg>` and `<rect>` elements with prefix svg (the svg namespace `http://www.w3.org/2000/svg`). However, if the file is saved as a HTML file, all the elements will be treated as HTML tags, as ":" can be used in element names, elements `<svg:svg>` and `<svg:rect>` will simply be considered as two elements with names svg:svg and svg:rect. This explains why SVG code cannot be parsed properly in HTML files.

## 4.5 How to make inline SVG work in HTML?

If you fully understand why inline SVG does not work in HTML files from last section, you may presume that the only issue we need to conquer is to manage to make web browsers consider tags such like `<svg:svg>` in HTML files to be XML elements but not HTML elements.

Before I discuss how to solve this problem, I would like to say that the XML parser is a function of web browsers, the parser exists and can be invoked by all kinds of web pages (no matter whether it is called from HTML or XML).

The strategy of solving the issue is creating all the SVG elements by using Javascript DOM function `document.createElementNS(namespace, tagName)`. This function will create an XML element with a specific namespace, and of course, the elements created by this function will be recognized as XML elements by web browsers.

For example, the following code has the same DOM structure as the code in

Example 4-4 and can be executed after saving as a HTML file:

```html
<html>
    <head>
        <script type="text/javascript">
            function loadSVG(){
                var oSvg = document.createElementNS("http://www.w3.org/2000/svg","svg:svg");
                        oSvg.setAttribute('viewBox','0 0 800 800');
                        oSvg.setAttribute('width',800);
                        oSvg.setAttribute('height',800);
                var oRect = document.createElementNS("http://www.w3.org/2000/svg","svg:rect");
                        oRect.setAttribute('x',1);
                        oRect.setAttribute('y',1);
                        oRect.setAttribute('width',1198);
                        oRect.setAttribute('height',398);
                        oRect.setAttribute('fill','red');
                        oRect.setAttribute('stroke','blue');
                        oRect.setAttribute('stroke-width',2);
                oSvg.appendChild(oRect);
                document.body.appendChild(oSvg);
            }
        </script>
    </head>
    <body onload="loadSVG()">
    </body>
</html>
```

**Example 4-5. SVG code created by DOM method**

According to this, we can create a generic function which first reads in all the SVG elements and then use `document.createElementNS()` to recreate these elements, which makes inline SVG possible in HTML file. An example is as follow:

```html
<html>
    <head>
        <script type="text/javascript">
            function recreateSVG(){
                //Read in all SVG elements
                var oAllSvgElements = getAllSvgElements();
                //Recreate all SVG elements
                var oAllRecreatedSvgElements =
                recreateSvgElements(oAllSvgElements);
                //Replace the previous SVG elements by the new ones
                refreshSvgElements(oAllRecreatedSvgElements);
            }
        </script>
    </head>
    <body onload="recreateSVG ()">
        <svg:svg viewBox="0 0 800 800" width="800" height="800"
                xmlns:svg="http://www.w3.org/2000/svg">
            <svg:rect x="1" y="1" width="1198" height="398"
                    fill="red" stroke="blue" stroke-width="2" />
        </svg:svg>
    </body>
</html>
```

**Example 4-6. Generic Recreating SVG Function**

Fortunately, this generic approach has already been implemented in Backbase Client Framework. Client Framework provides a binding language called Tag

Definition Language (TDL). TDL is originally designed for extending BTL (Backbase Tag Library) UI widgets, adding skins, new language implementations, as well as creating new widgets or XML languages. It's an easy and powerful mechanism for developers to create new widgets in their own namespace. (For syntax and more information about TDL, please look in the *Widget Development Guide.pdf* of Backbase).

The mechanism of Backbase Client Framework is that when the web document is fully loaded, the framework will first read in all the elements within the Backbase `<script>` tag, then the Backbase Client Framework engine will use Backbase bindings (which is defined by TDL) to process, parse or recreate every element in different approaches according to the namespace it belongs to. This process is very similar to the process of recreating the SVG elements that I explained above.

Take the advantage of TDL, every SVG element can be considered as a simple widget tag, so all the SVG tags can be defined before hand in TDL, a simple example is as follow:

```
<d:tdl xmlns:d="http://www.backbase.com/2006/tdl">
<d:namespace name="http://www.w3.org/2000/svg">
<d:element name="svg">
        <d:attribute name="viewBox">
            <d:mapper type="text/javascript"><![CDATA[
                this.viewNode.setAttribute(name, value);
            ]]></d:mapper>
        </d:attribute>
        <d:attribute name="width">
            <d:mapper type="text/javascript"><![CDATA[
                this.viewNode.setAttribute(name, value);
            ]]></d:mapper>
        </d:attribute>
        //…… More attribute definition
</d:element>
</d:namespace>
</d:tdl>
```

**Example 4-7. SVG elements defined by TDL**

Since all the SVG elements can be defined in the same way, native inline SVG for HTML files is implemented by using TDL in this project.

# 5. How to implement SVG in Internet Explorer ?

In chapter 4, all the issues are discussed based on standard browsers, from this chapter on, I will bring you into the core of this thesis, how to implement SVG in Internet Explorer.

SVG is not natively supported by Microsoft Internet Explorer. In order to view SVG graphics, third party plug-ins (e.g. Adobe SVG Viewer) must be installed into IE. Instead of SVG, Internet Explorer has its own method to support vector graphics – VML. In most of the cases, VML can create the same or similar graphics or effects as SVG.

The process of implementing SVG in IE is actually the process of translating SVG into VML. For more information about why I choose VML to be the counterpart of SVG, please refer to my research project report article "*How To Select an Implementation for Vector Graphics Across Browsers*".

## 5.1 Feasibility study

Both SVG and VML are languages in XML, so they share some common characteristics such as DOM accessibility and DOM manipulation. They are also languages for 2D vector graphics, so they both use XML (a very elegant way) to present nice vector graphics.

However, SVG and VML are created for different purpose in different times. They have different syntaxes and functionalities.

The aim of this project is not translating all SVG elements into VML, there are several reasons for this. First, it's just impossible. As I have said above, SVG and VML have different functionalities, some of the functionalities are exclusive. For example, in SVG we can define an arrow with a customized arrowhead as follows, which is impossible in VML. Second, performance consideration. SVG and VML have different syntaxes, so you may have to provide different parameters or attribute values in order to create a same graphic. The calculating from SVG into VML can be very expensive if the syntaxes are too different in which case it is not worth it. Lastly, some of the SVG elements are only useful in document structure perspective, but considered to be useless in visual perspective. For example, `<svg:desc>` and `<svg:title>` element, when the SVG document fragment is rendered as SVG on visual media, `<svg:desc>` and `<svg:title>` elements are not rendered as part of the graphics, so the only usage of these elements is to provide document structure information.

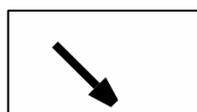

**Figure 5-1. SVG customized arrowhead**

For these reasons above, I first created a list which indicates which SVG elements are going to be implemented and their VML counterparts. You can find the full list in the appendix of this report. From the list we can see that most of the fundamental SVG elements have their corresponding VML elements, with these elements, most of the charts as well as shapes can be created, in other words, it's feasible to use VML to be the shadow content of SVG.

## 5.2 Requirement Analysis on Backbase Client Framework

The following snippet contains the possible code in a Backbase page:

```
<html>
    <head>
        <script type="text/javascript"
          src="Backbase_4_2_0/engine/boot.js"></script></head>
    <body>
        <script xmlns="http://www.w3.org/1999/xhtml"
            type="application/backbase+xml">
            <xi:include href="Backbase_4_2_0/bindings/config.xml" />
            <d:namespace name="http://myns">
                <d:element name="canvas">
                    <d:attribute name="width"/>
                    <d:template type="image/svg+xml">
                        <svg:svg viewBox="0 0 100 100">  //SVG in d:template
                            <d:content />
                        </svg:svg>
                    </d:template>
                </d:element>
            </d:namespace>
            <my:canvas width="100">
                //SVG in d:content
                <svg:rect x="10" y="10" width="398" height="198"/>
            </my:canvas>
            <svg:svg viewBox="0 0 100 100">  //SVG in document
                ……
            </svg:svg>
        </script>
    </body>
</html>
```

**Example 5-1. Typical Backbase Client Framework Code Format**

From the code above, we can see that SVG code can appear in three places: d:template, d:content and document.

Considering the time of this project, only SVG in document has been implemented. The other two are left to be future work.

## 5.3 First Cross-browser SVG Example

In this section, I will implement the first SVG example. The implementation is finished by using Backbase TDL. The purpose of this example is to show a SVG rectangle in all of the browsers.

The SVG syntax of drawing a rectangle is as follow:

```
<svg:svg viewBox="0 0 800 800" width="800" height="800"
        xmlns:svg="http://www.w3.org/2000/svg">
    <svg:rect x="1" y="1" width="1198" height="398" fill="red"
            stroke="blue" stroke-width="2" />
</svg:svg>
```

**Example 5-2. SVG syntax of drawing a rectangle**

We need two SVG elements "svg:svg" and "svg:rect" to draw a rectangle in the browser.

First, let's look at svg:svg element. From the list in the appendix we can see that the counterpart VML element for svg:svg element is vml:group element, so the first part of code should be written as follow:

```
<d:element name="svg">
  <d:template type="text/javascript"><![CDATA[
    if (bb.browser.ie) {
        document.createElement("vml:group");
    } else {
        document.createElementNS("http://www.w3.org/2000/svg", "svg:svg");
    }
  ]]></d:template>
</d:element>
```

**Example 5-3. TDL of defining svg:svg element**

"bb.browser.ie" is a constant in Backbase Client Framework indicating whether the current browser is Internet Explorer. If the current browser is IE, a VML element vml:group will be created by using document.createElement("vml:group") into the document, otherwise, a SVG element with the same name will be created by using document.createElementNS("http://www.w3.org/2000/svg", "svg:svg").

Now we can use Backbase Debugger to have a look at what the document structure looks like in different browsers:

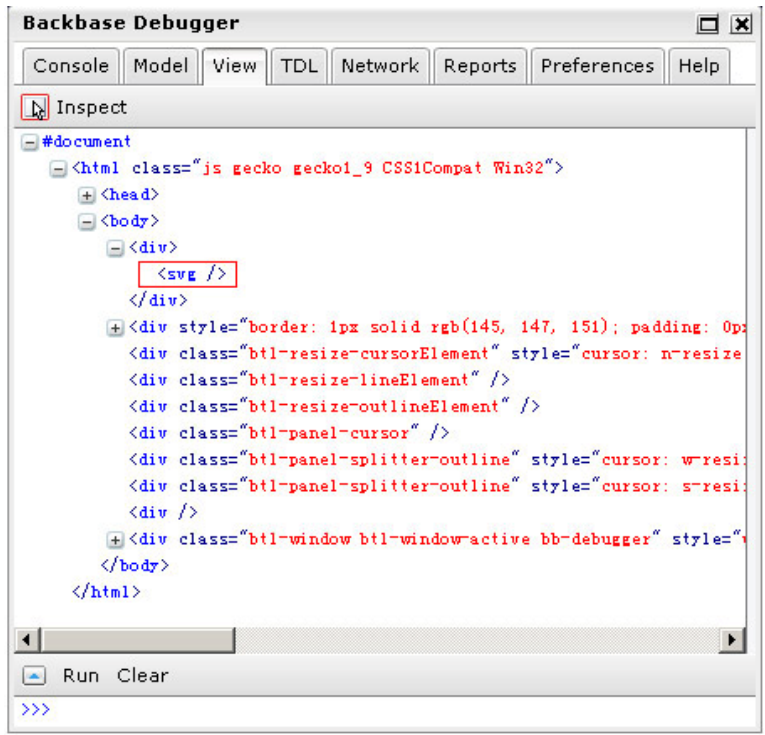

**Figure 5-2. Document Structure in Firefox**

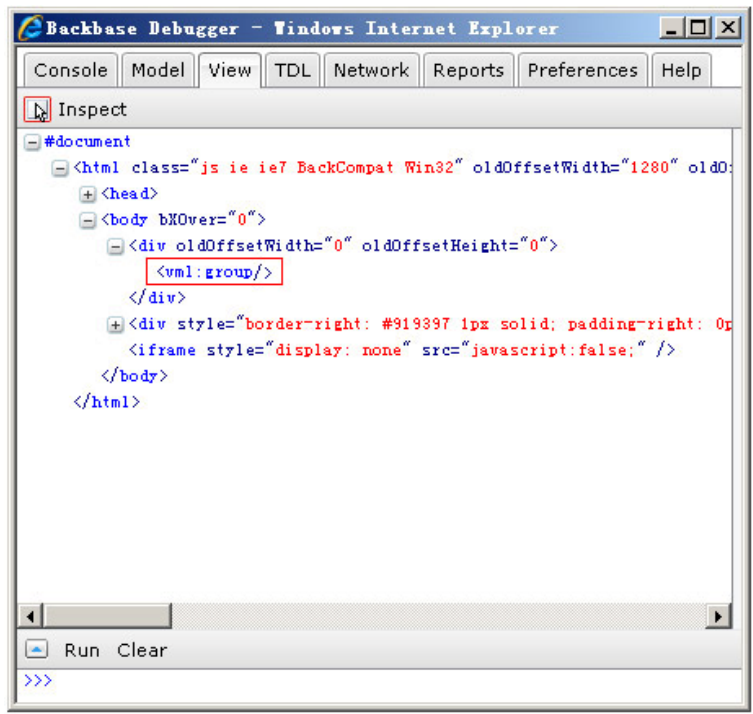

**Figure 5-3. Document Structure in IE**

From the document structure above, we can see that, we have generated different object models in different browsers. You may notice that the attributes for the elements are still missing, so the next step is to set the attribute values.

For svg:svg element, there are three possible attributes: viewBox, width and height. The following table shows their counterpart VML attributes:

| svg:svg attributes | Vml:group attributes |
|---|---|
| viewBox | coordOrigin, coordSize |
| Width | style:width |
| height | style:height |

**Table 5-1. SVG and VML attributes**

Base on the table above, I extended the previous code into following:

```
<d:element name="svg">
    <d:template type="text/javascript"><![CDATA[
        if (bb.browser.ie) {
            document.createElement("vml:group");
        } else {
            document.createElementNS("http://www.w3.org/2000/svg",
                                     "svg:svg");
        }
    ]]></d:template>
    <d:attribute name="viewBox">
        <d:mapper type="text/javascript"><![CDATA[
            if (bb.browser.ie) {
                var aValue = value.split(/\s+/);
                this.viewNode.coordOrigin = aValue[0] + ',' + aValue[1];
                this.viewNode.coordSize = aValue[2] + ',' + aValue[3];
            } else {
                this.viewNode.setAttribute(name, value);
            }
        ]]></d:mapper>
    </d:attribute>
    <d:attribute name="width">
        <d:mapper type="text/javascript"><![CDATA[
            if (bb.browser.ie) {
                this.viewNode.style[name] = value;
            } else {
                this.viewNode.setAttribute(name, value);
            }
        ]]></d:mapper>
    </d:attribute>
    <d:attribute name="height">
        <d:mapper type="text/javascript"><![CDATA[
            if (bb.browser.ie) {
                this.viewNode.style[name] = value;
            } else {
                this.viewNode.setAttribute(name, value);
            }
        ]]></d:mapper>
    </d:attribute>
</d:element>
```

**Example 5-4. TDL of defining svg:svg element**

Note that I use Javascript Regular Expression to assign the viewBox value to coordOrigin and coordSize.

Then let's look at the document structure again:

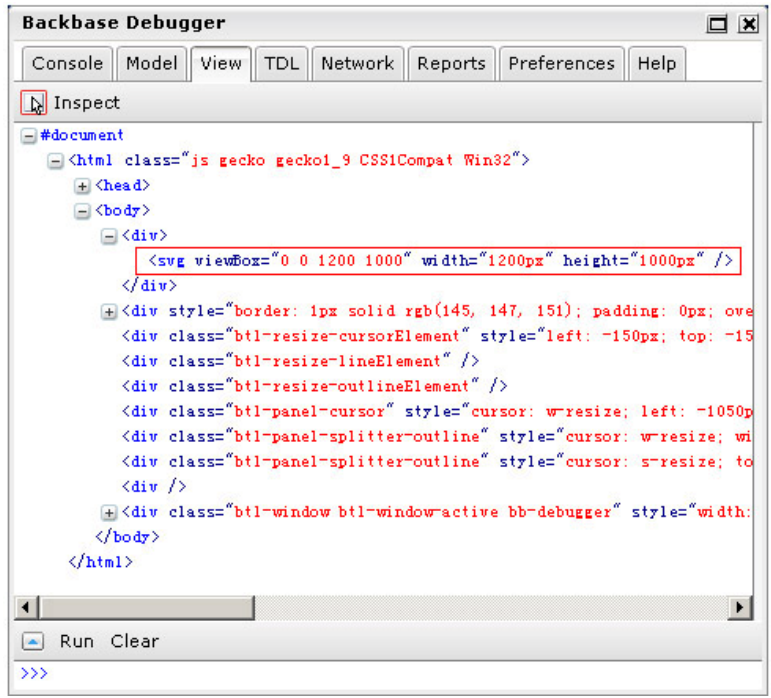

**Figure 5-4. Document Structure in FF**

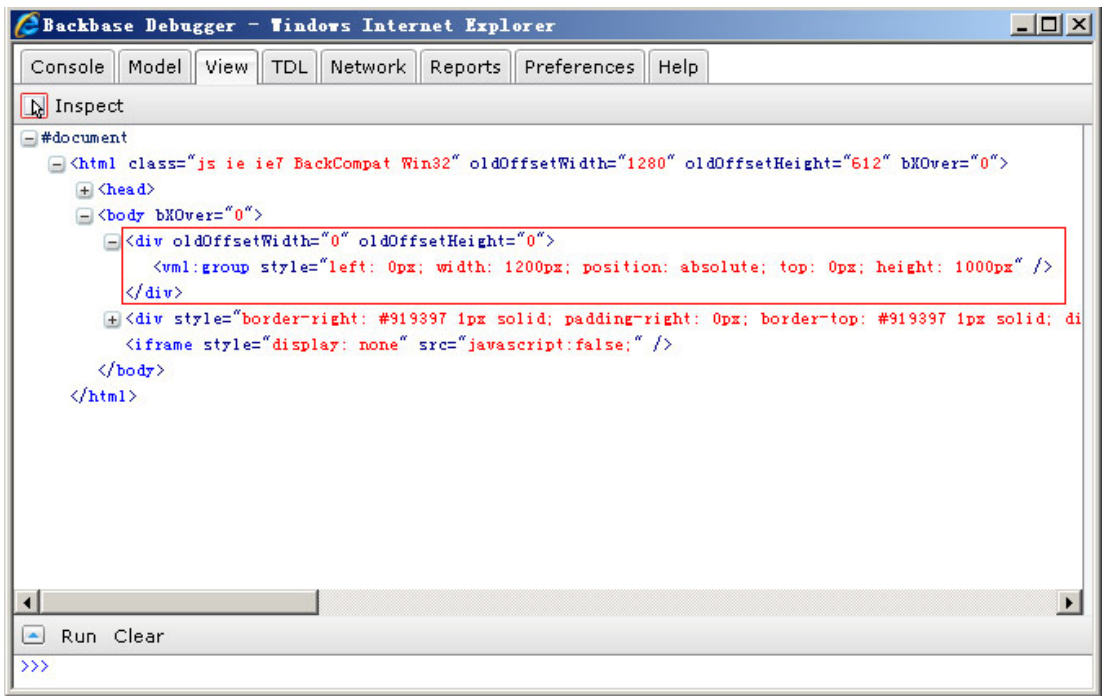

**Figure 5-5. Document Structure in IE**

Please note that Backbase debugger in IE cannot show coordOrigin and coordSize attributes.

So far, the svg:svg element has been implemented. We can use the same approach to implement the svg:rect element, the code is as follow:

```
<d:element name="rect">
    <d:template type="text/javascript"><![CDATA[
        if (bb.browser.ie) {
            document.createElement("vml:roundrect");
        } else {
            document.createElementNS("http://www.w3.org/2000/svg",
        "svg:rect");
        }
    ]]></d:template>
    <d:attribute name="x">
        <d:mapper type="text/javascript"><![CDATA[
            if (bb.browser.ie) {
                this.viewNode.style.left = value;
            } else {
                this.viewNode.setAttribute(name, value);
            }
        ]]></d:mapper>
    </d:attribute>
    <d:attribute name="y">
        <d:mapper type="text/javascript"><![CDATA[
            if (bb.browser.ie) {
                this.viewNode.style.top = value;
            } else {
                this.viewNode.setAttribute(name, value);
            }
        ]]></d:mapper>
    </d:attribute>
    <d:attribute name="width">
        <d:mapper type="text/javascript"><![CDATA[
            if (bb.browser.ie) {
                this.viewNode.style.width = value;
            } else {
                this.viewNode.setAttribute(name, value);
            }
        ]]></d:mapper>
    </d:attribute>
    <d:attribute name="height">
        <d:mapper type="text/javascript"><![CDATA[
            if (bb.browser.ie) {
                this.viewNode.style.height = value;
            } else {
                this.viewNode.setAttribute(name, value);
            }
        ]]></d:mapper>
    </d:attribute>
    <d:attribute name="rx">
        <d:mapper type="text/javascript"><![CDATA[
            if (bb.browser.ie) {
                this.viewNode.arcsize = parseInt(value) /
(parseInt(this.viewNode.style.width) / 2 ) ;
            } else {
                this.viewNode.setAttribute(name, value);
            }
        ]]></d:mapper>
    </d:attribute>
    <d:attribute name="ry">
        <d:mapper type="text/javascript"><![CDATA[
            if (bb.browser.ie) {
                this.viewNode.arcsize = parseInt(value) /
( parseInt(this.viewNode.style.height) / 2 ) ;
            } else {
                this.viewNode.setAttribute(name, value);
            }
        ]]></d:mapper>
    </d:attribute>
```

**Example 5-5. TDL of defining svg:rect element**

Now both of the two elements have been implemented. The result is as follow:

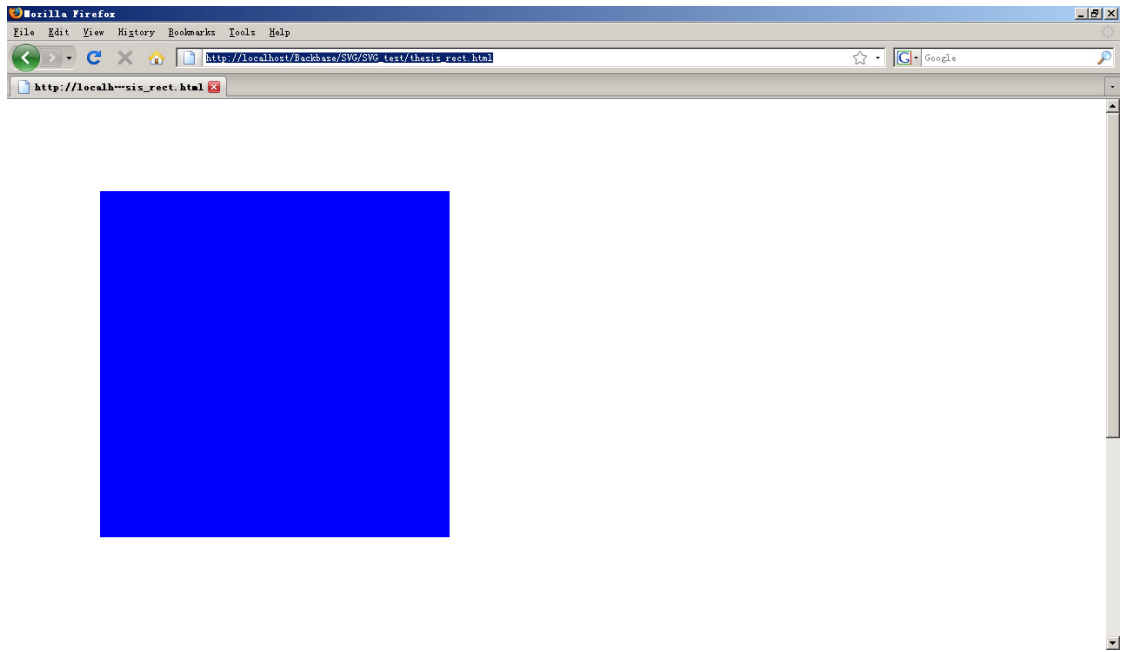

**Figure 5-6. First Example in Firefox**

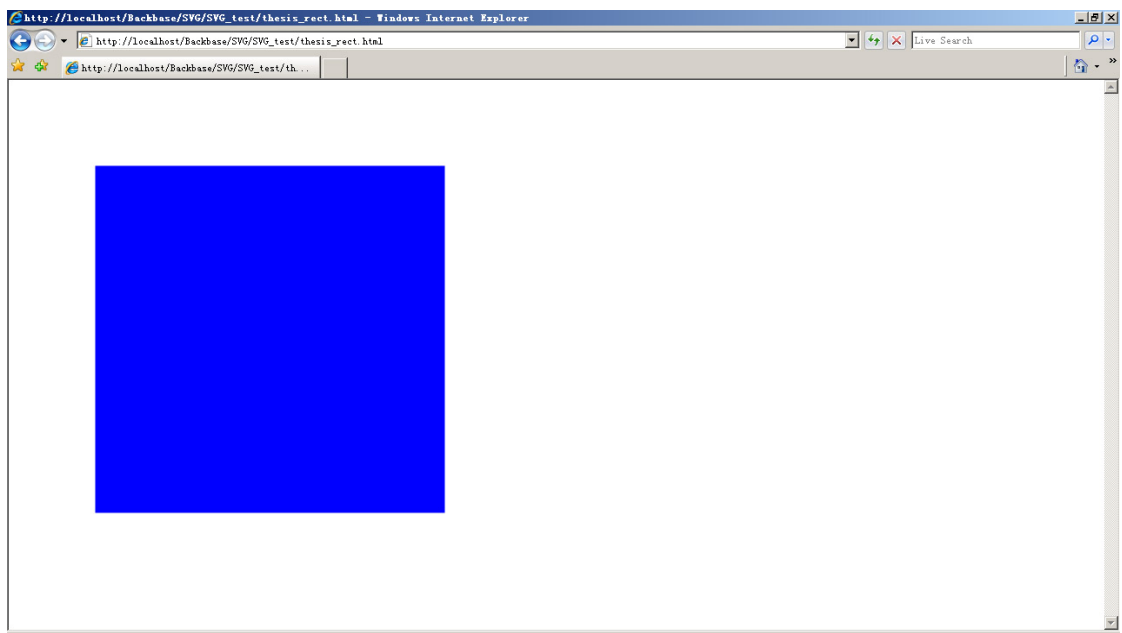

**Figure 5-7. First Example in IE**

In this chapter, I explained how to translate SVG into VML by using Backbase TDL so that SVG can be shown in all browsers. In order to let you understand how it works more easily, I pasted all the related codes. From the next chapter on, I will only show the key code so to save space.

## 5.4 DOM and DOM Tree

### 5.4.1 DOM – Document Object Model

What is the Document Oject Model? The Document Object Model is a platform-
and language-neutral interface that will allow programs and scripts to
dynamically access and update the content, structure and style of documents.
The document can be further processed and the results of that processing can
be incorporated back into the presented page.

### 5.4.2 SVG DOM Tree and VML DOM Tree

In last section, a SVG document was translated into a VML document. A SVG
document can also be considered as a tree whose root is the SVG root element
"svg:svg", the leaves are other descendant nodes.

Similarly, the counterpart VML/HTML elements also build up a new tree.

In order to make it simple, in the following chapters, SVG document is called
"SVG DOM Tree", the counterpart VML/HTML document is called "VML DOM Tree".
The procedure of translating SVG elements to their counterpart VML/HTML is
called "mapping".

# 6. Implementation of SVG elements

In this and the next chapter I will explain how I implemented all the SVG elements and attributes in this project.

## 6.1 Implementing SVG in Object-oriented Programming

In SVG, different elements share some same attribute. For example, attribute "id" is an common attribute for all elements. Another example is attribute "transform" - all the shape elements as well as group element have this attribute. In order to simplify the implementation, I take the advantage of Object-oriented programming (OOP) feature in TDL to create some common class, which makes the code more elegant and easier to maintain.

There are totally four parent abstract classes: *SVGElement*, *SVGStylable*, *SVGTransformable* and *referenceElement*.

*SVGElement* defines the common attribute "id". The template of *SVGElement* returns the viewNode and viewGate for all the elements. viewNode is the root node of the visual structure that represents the element; viewGate is the node were you will append new presentation child-nodes if required. (For more information of viewNode and viewGate, please refer to "*Application Development Guide*" for the definition of viewNode) Therefore, all the child elements should inherit from this class.

*SVGStyleable* defines the common attribute which related with style, such as "style", "color", "fill", "stroke" and so on. All the shape elements, "text", "textPath" as well as "path" and "foreignObject" elements inherit from this class.

*SVGTransformable* is defined for the elements that can be transformed, so all the shape elements, "path" element and "foreignObject" element inherit from this class.

In SVG, some of the elements have attributes that reference to other elements, such like "use" and "textPath" elements. *referenceElement* is defined for these elements to preserve the reference elements.

The charts below show the inheritance of all the elements.

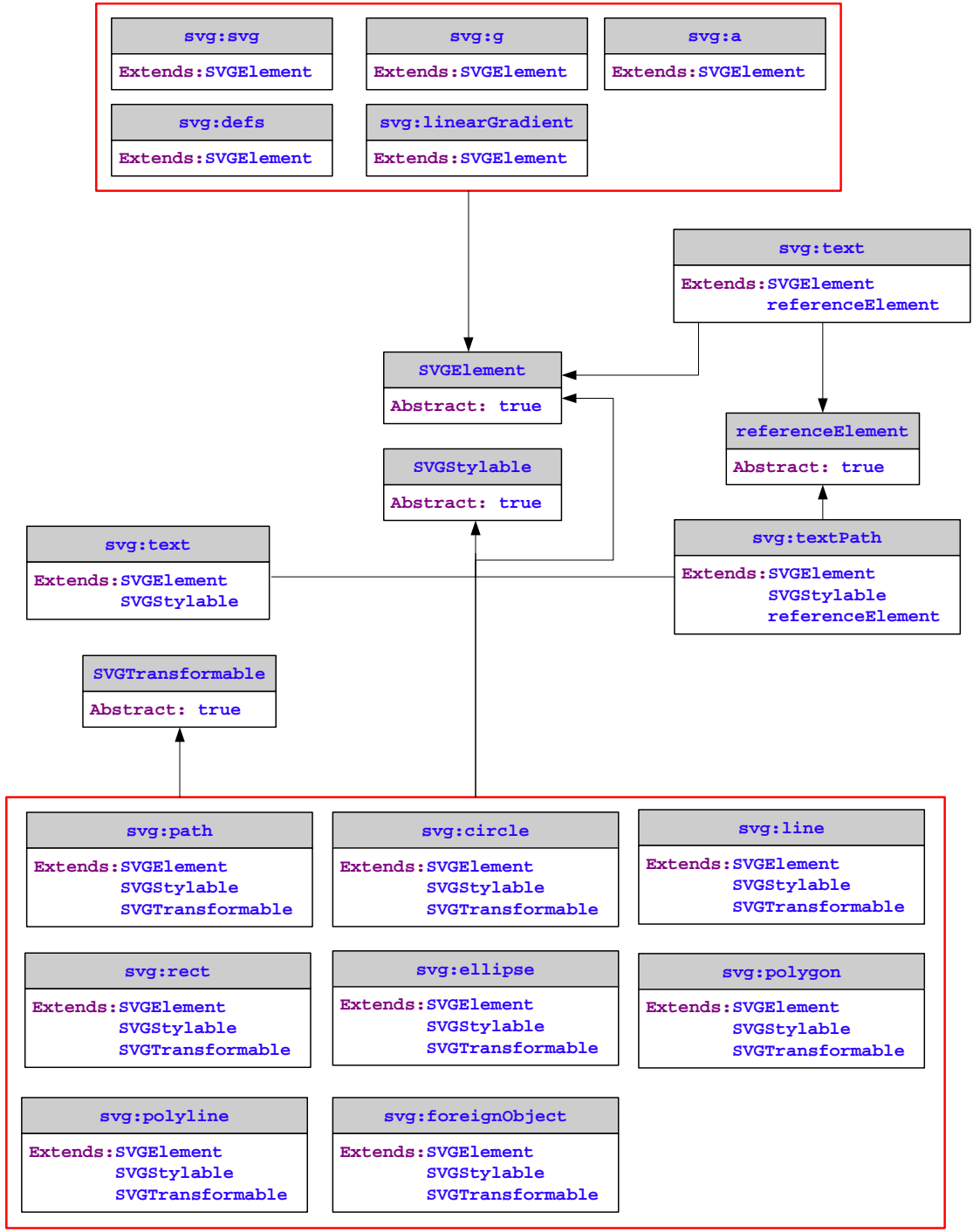

**Chart 6-1. Inheritance of All elements**

## 6.2 Implementation of Structure Elements

### 6.2.1 The <svg> element

The equivalent VML element of "svg" element is "vml:group". For more information about the implementation, please refer to the last chapter.

### 6.2.2 The <g> element

The "g" element is a container element for grouping together related graphics elements. Once grouped you can transform the whole group of shapes or set the fill or stroke attributes for all the shapes. That is to say, all the sub-elements will inherit the features from the "g" element. The equivalent VML element is "vml:group".

I created a function `applyToChildNodes(thisNode, attrName, attrValue)` to do the inheriting. For example, we can use

```
<d:attribute name="fill">
    <d:mapper type="text/javascript"><![CDATA[
        if (bb.browser.ie) {
            this.overwriteChildNodesAttributes(this, name, value);
        }
    ]]></d:mapper>
</d:attribute>
```

**Example 6-1. Apply features to descendant nodes**

to have each sub-element of "g" element to inherit the value of its "fill" attribute. I used this approach to implement other attributes of "g" element, such as "stroke", "stroke-width", "opacity" and "transform".

## 6.3 Implementation of Shape Elements

### 6.3.1 The <rect> element

The SVG "rect" element has attributes "x", "y", "width", and "height", which map directly to VML "left", "top", "width", and "height" attributes respectively for the "vml:rect" element.

### 6.3.2 The <circle> and <ellipse> element

The equivalent VML element of SVG "circle" and "ellipse" elements is "vml:oval", but with some attribute differences. In SVG "circle" element, the x coordinate of the center of the circle is represented by the "cx" attribute, the y coordinate of the center is indicated by the "cy" attribute, and "r" refers to the radius. In VML, the "vml:oval" element uses "top", "left", "width", and "height" attributes to define an oval. The Figure below compares the attributes of SVG "circle" element and "vml:oval" element:

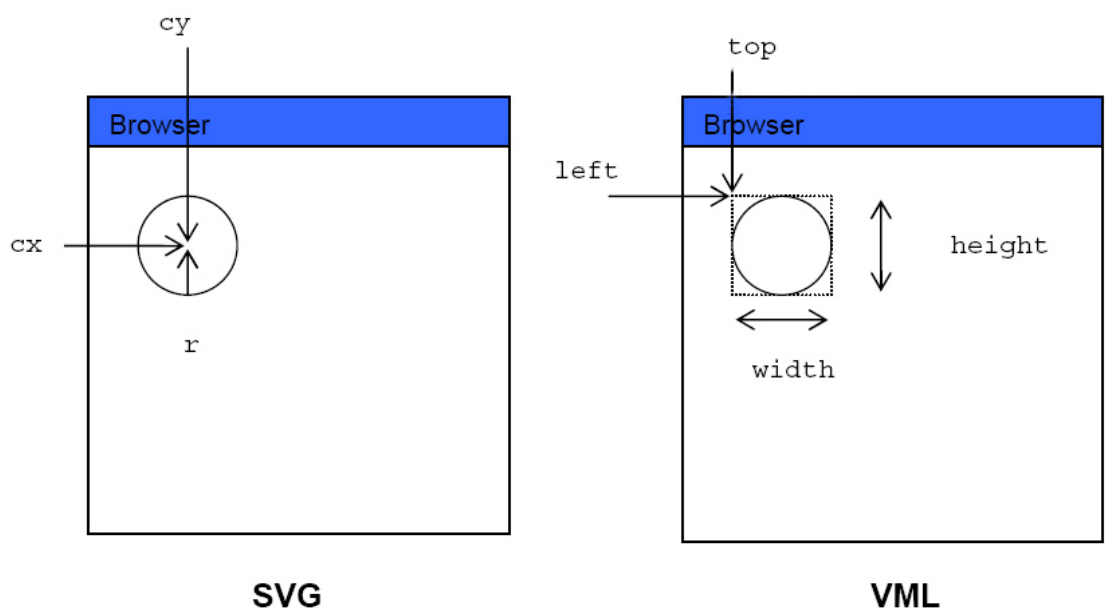

**SVG**             **VML**

**Figure 6-1. Attributes of svg:circle and vml:oval**

The calculation is straightforward as follow:

$$width = r * 2;$$
$$height = r * 2;$$
$$top = cy - r;$$
$$left = cx - r;$$

**Formula 6-1. Calculation from svg:circle attributes into vml:oval attributes**

The only difference between SVG "ellipse" and "circle" elements are the "rx" and "ry" attributes that the "ellipse" element has. The calculations is as follow:

$$width = r * 2;$$
$$height = r * 2;$$
$$top = cy - rx;$$
$$left = cx - ry;$$

**Formula 6-2. Calculation from svg:ellipse attributes into vml:oval attributes**

### 6.3.3 The <polyline> <polygon> and <line> elements

They all map to "vml:shape" element. "vml:shape" has attribute "path" containing the coordinates and commands that define the path.

SVG "polyline" and "polygon" elements both contain the attribute "points" representing a series of points between which the straight lines are drawn. SVG "line" element has attributes "x1", "y1", "x2", "y2" which determine the start point and the end point of the line.

For SVG "polyline" and "polygon" elements, I first split the value of attribute "points" into a series of point coordinates; then create a path string which combines the coordinates and "m", "l", "x" and "e" commands; at last assign the string to "path" attribute. The difference between SVG "polyline" and "polygon" is SVG "polygon" will close the shape automatically, so the path string for "polygon" should be end with "x" and "e" commands while the path string for "polyline" should only be end with "e" command.

For example, `<svg:polyline points="0,0 100,100 200,200"/>` should be translated into `<vml:shape path="m 0,0 l 100,100 l 200,200 e"/>` while `<svg:polygon points="0,0 100,100 200,200"/>` should be translated into `<vml:shape path="m 0,0 l 100,100 l 200,200 x e"/>`.

The implementation of an SVG "line" element is much simpler. The path string for an SVG "line" element is a combination of the start point and the end point with the commands. For example, `<svg:line x1="0" y1="0" x2="100" y2="200"/>` should be translated into `<vml:shape path="m 0,0 l 100,200 e"/>`.

## 6.4 Implementation of <path> Element

The SVG "path" element is more powerful than the shape elements above. With a SVG "path" element, you can draw almost all kinds of shapes. Its most important attribute is "d" attribute, which is the definition of the outline of a shape. The "d" attribute contains the *moveto*, *line*, *curve* (both cubic and quadratic Béziers), *arc* and *closepath* commands. It also maps to "vml:shape" element.

The "d" attribute of SVG "path" element is very similar to "vml:shape" 's "path" attribute. The commands are also very similar. The following table is a complete list of commands in SVG and their VML counterparts as well as their implementation status:

| Commands in SVG | VML Counterparts | Implementation Status |
|---|---|---|
| M, m | m | Yes |
| Z, z | x, e or e | Yes |
| L, l | l | Yes |
| H, h | l | Yes |
| V, v | l | Yes |
| C, c | c | Yes |
| S, s | Not found | Not going to implement |
| Q, q | Not found | Not going to implement |
| T, t | Not found | Not going to implement |
| A, a | Not found | Further work |

**Table 6-1. Commands Implementing Status**

The upper case commands in SVG operate based on absolute coordinates while lower case commands based on relative coordinates. On the other hand, all the commands in VML are in lower case, all of which operate on absolute coordinates. For example, <svg:path d="m 100,100 M 200,200 z"/> will be translated into <vml:shape path="m 100,100 m 200,200"/> while <svg:path d="m 100,100 m 200,200 z"/> will be translated into <vml:shape path="m 100,100 m300,300"/>.

Besides, in SVG, supposing the current position is (x,y), then we have the following conclusion:
1. command "H sx" equals "L sx+x,y"
2. command "h sx" equals "l sx,0"
3. command "V sy" equals "L x, sy+y"
3. command "v sy" equals "l 0, sy"
This is why the counterpart VML command of SVG "H", "h" and "V", "v" commands is "l".

## 6.5 Implementation of <foreignObject> Element

The SVG "foreignObject" element allows for inclusion of a foreign namespace which has its graphical content drawn by a different user agent. The included foreign graphical content is subject to SVG transformations and compositing. A very fascinating application is to use "foreignObject" do some fancy transformations on HTML contents. The following figure is an example:

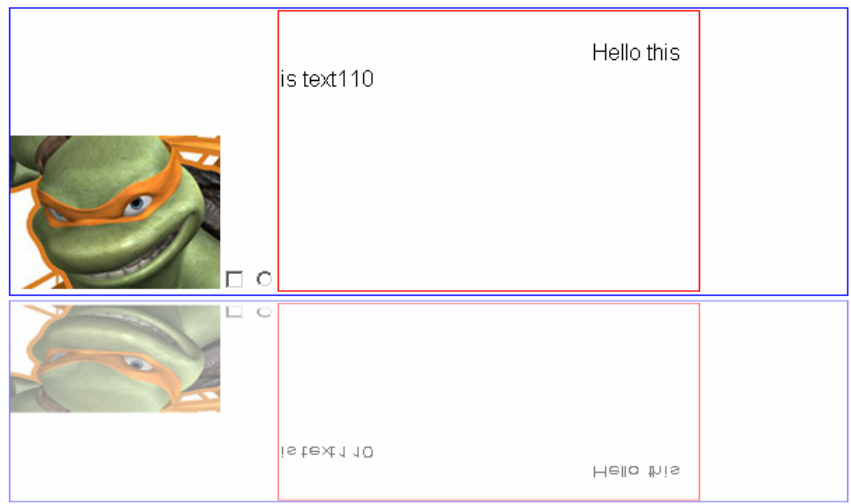

**Figure 6-2. Example of foreignObject**

I use "vml:textBox" element to implement SVG "foreignObject" element. The attributes "x", "y" and "width" , "height" values of SVG "foreignObject" element can be directly mapped into "vml:textBox" element's "style:left", "style:top" and "style:width", "style:height" values. The "opacity" attribute is implemented by using "CSS Alpha Filter". The only difference is that before assign the opacity value to "CSS Alpha Filter" the value must be multiplied by 100 because the opacity value in "CSS Alpha Filter" ranges from 0 to 100 while in SVG it ranges from 0 to 1.

## 6.6 Implementation of <text> Element

The SVG "text" element is also implemented by "vml:textBox" element. The "x" attribute of "svg:text" can be mapped directly into "style:left" of "vml:textBox" element, whereas an extra calculation is needed when mapping the "y" attribute to "style:top" of "vml:textBox":

$$style:top = \ x - fontSize;$$

**Formula 6-3. Mapping "y" attribute to "style:top"**

## 6.7 Implementation of reference Elements

Reference elements are the elements whose attributes contain references to other elements. For example,

```
<svg:svg viewBox="0 0 1000 1500" width="1000px" height="1500px">
    <svg:defs>
        <svg:rect id="MyRect" x="100" y="200"
                      width="600" height="200" fill="blue" stroke="red"/>
    </svg:defs>
    <svg:use xlink:href="#MyRect"/>
</svg:svg>
```

**Example 6-2. Reference Element example**

attribute "xlink:href" of "svg:use" element refers to element "MyRect".

### 6.7.1 The <defs> and <use> elements

The SVG "defs" element is a container element for referenced elements. Elements that are descendants of a "defs" element are not rendered directly, in other words, if a "defs" element is not referenced, its content is invisible. Therefore, I map the SVG "defs" element to "html:div" element, and then set the visibility of "html:div" element to be invisible.

The SVG "use" element has optional attributes x, y, width and height which are used to map the graphical contents of the referenced element onto a rectangular region within the current coordinate system. The implementation of this element is like follows:

1. Map the "svg:use" element to the "html:div" element.
2. Copy all the descendant elements of the referenced "svg:defs" element and then append these descendant elements to "html:div" element.
3. Map "x", "y", "width" and "height" attributes of "svg:use" element to "style:left", "style:top", "style:width" and "style:height" attributes of "html:div" element.

### 6.7.2 The <textPath> element

In addition to text drawn in a straight line, the SVG "textPath" element gives SVG the power to place text along the shape of a SVG "path" element.

VML also has the "textPath" ability. The figure below shows how these two different languages generate text paths and their relationship.

```
<svg:svg width="1000" height="300" viewBox="0 0 1000 300">
    <svg:defs>
        <svg:path id="MyPath" d="M 100 200
                                 C 200 100 300 0 400 100
                                 C 500 200 600 300 700 200
                                 C 800 100 900 100 900 100" />
    </svg:defs>
    <svg:text font-family="Verdana" font-size="40px">
        <svg:textPath xlink:href="#MyPath">
            We go up, then we go down
        </svg:textPath>
    </svg:text>
</svg:svg>

<vml:shape path="m 100 200
                 c 200 100 300 0 400 100
                 c 500 200 600 300 700 200
                 c 800 100 900 100 900 100">
    <vml:path textpathok="t"/>
    <vml:textpath style="FONT-SIZE:40;FONT-FAMILY:Verdana"
                  on="t" string="We go up, then we go down"/>
</vml:shape>
```

**Figure 6-3. textPath code in SVG and VML**

From the figure above we can see the implementation can be divided into the following steps:

Step1. Find the reference "svg:path" element whose value of "id" attribute is "MyPath" and translated it into a "vml:shape" element.

Step2. Create a "vml:path" element with its attribute "textpathok" being "t" and a "vml:textPath" element with its attribute "on" being "t", then append these two elements to the "vml:shape" element in step1.

Step3. Find the parent node of "svg:textPath" element. The parent node should be a "svg:text" element and then map its attributes to "style" attribute of "vml:textpath" element.

Step4. Get the content of "svg:textPath" element, then map the content to the "string" attribute of "vml:textpath" element.

### 6.7.3 The <linearGradient> and <stop> elements

Gradient is a smooth transition of one color to another. There are two types of gradient in SVG: linear and radial. Predefined shapes can use gradients as fill colors. Linear gradients can be horizontal, vertical, or diagonal. I only implemented the horizontal and vertical linear gradients of up to two colors with offsets of 0% and 100% to indicate smooth transition.

SVG "linearGradient" and "stop" elements are actually implemented in the common attribute - "style", for more information please refer to section 7.2.

## 6.8 Implementation of <a> Element

The implementation of SVG "a" element is the simplest. I first map the "svg:a" element to a "html:a" element and then map attribute "xlink:href" of "svg:a" element to attribute "href" of "html:a" element.

# 7. Abstract classes - implementation of SVG common attributes

## 7.1 The abstract class "SVGElement"

Abstract class "SVGElement" is the most fundamental class. The implementation of every SVG element should inherit from this class.

The d:template part of this class is used to create either SVG or their counterpart elements according to the browsers.

Attribute "id" is implemented in the d:attribute part. "id" is the only attribute that every SVG element can have. The implementation is very straightforward - just map the attribute to the corresponding VML element's "id" attribute.

## 7.2 The abstract class "SVGStylable"

This class is defined by several d:attributes, d:properties, one d:method and one d:handler. This class is used to map SVG presentation attributes.

### 7.2.1 d:handler "DOMNodeInsertedIntoDocument"

Let's first look at the d:handler "DOMNodeInsertedIntoDocument". "handler" (also called "event handler") is a term in Backbase Client Framework, which can be considered as a function that will be triggered only when a certain event happens. Take d:handler "DOMNodeInsertedIntoDocument" for an example, the content of this handler will only be executed when "the DOM tree is inserted into the document tree" happens.

### 7.2.2 "stroke" related attributes

When mapping the attributes that are related with "stroke", d:property "strokeElement" will be called. The "getter" of "strokeElement" will create a "vml:stroke" node which will be appended as a child node to the context node of the VML DOM tree. Then the "mappers" of d:attributes will map the stroke features to "vml:stroke". The following table shows the mapping relationship:

| Attributes in SVG | Counterpart Attributes of <vml:stroke> |
|---|---|
| stroke | color |
| stroke-width | Weight |
| stroke-linecap | endcap |
| stroke-linejoin | joinstyle |
| stroke-miterlimit | miterlimit |
| stroke-opacity | opacity |

**Table 7-1. Attributes in SVG and VML**

### 7.2.3 "fill" attribute

Mapping the "fill" attribute is very similar to mapping attributes that are related with "stroke". The mapping will first call d:property "fillElement" which will create a "vml:fill" element; then this element will be appended as a child element to the context node of the VML DOM tree; at last the "mapper" of d:attribute "fill" will map the fill features to "vml:fill".

The only difference of mapping "fill" attribute and "stroke" attribute is that the value of "fill" attribute can either be a solid color or a gradient one. The mapping method mentioned above is only applicable to the solid fill color. This project supports horizontal and vertical linear gradients of up to two colors with offsets of 0% and 100%.

When the fill color is a gradient one, d:method "setGradientFill" will be called. This method will first locate the "svg:linearGradient" element that the "fill" attribute refers to; then find the "svg:stop" elements who are the child nodes of "svg:linearGradient" element; at last, map the "stop-colors" in "svg:stop" element to "vml:fill" element.

## 7.3 The abstract class "referenceElement"

This class is specially designed for the SVG elements that were mentioned in Section 6.7.

Only one d:property "refElement" is defined in this class. The return value of this d:property is the copy of the reference node. For example, element `<svg: use xlink:href="#rect" />` refers to element `<svg:rect id="rect" />`, then d:property "refElement" in element `<svg: use xlink:href="#rect" />` is actually a copy of element `<svg:rect id="rect" />`. In the meantime, the changing of the d:property value will not affect the reference element.

d:property "refElement" simplify the implementation in Section 6.7 because:

1.  Supposing that element A has an attribute refers to element B. A copy of element B can always be stored in d:property "refElement" of element A.
2.  When re-mapping attributes of element A, we don't need to look for element B in the DOM tree once again but just retrieve it from d:property "refElement".

## 7.4 The abstract class "SVGTransformable"

All the transformable elements inherit from this class.

There are seven d:properties defined in this class:

1.  d:property "skewElement". This d:property first creates a "vml:skew" element and then append this element as a child node to the context element of the VML DOM tree. This d:property is only used by elements whose counterpart elements need "vml:skew" element to do the transformation.
2.  d:property "scaleXSize" and "scaleYSize". These two d:properties store the scale information. They will be used in SVG "path" element when both "scale" and "translate" appear in its "transform" attribute.
3.  d:property "originalX", "originalY", "originalWidth" and "originalHeight". These four d:properties store the original information of "x", "y", "width" and "height". They are used in DOM manipulation.

For more information about transformations, please refer to the next chapter.

# 8. Implementation of the "transform" attribute

Transformation is one of the most attractive features in SVG. All SVG graphic elements, "g" element as well as some other elements have a "transform" attribute. This attribute can be set and changed to various values in order to move and distort the element.

Simulating transformation effects is the most difficult and complex work in the whole implementation, that is why I wrote a separate chapter to explain how to do it.

## 8.1 Basic knowledge of transformations

### 8.1.1 Transformation types

The available types of transform are explained as follow:

- matrix(<a> <b> <c> <d> <e> <f>), which specifies a transformation in the form of a transformation matrix of six values. matrix(a,b,c,d,e,f) is equivalent to applying the transformation matrix **[a b c d e f]**.

- translate(<tx> [<ty>]), which specifies a translation by *tx* and *ty*. If *<ty>* is not provided, it is assumed to be zero.

- scale(<sx> [<sy>]), which specifies a scale operation by *sx* and *sy*. If *<sy>* is not provided, it is assumed to be equal to *<sx>*.

- rotate(<rotate-angle> [<cx> <cy>]), which specifies a rotation by <rotate-angle> degrees about a given point.
  If optional parameters <cx> and <cy> are not supplied, the rotate is about the origin of the current user coordinate system. The operation corresponds to the matrix **[cos(a) sin(a) -sin(a) cos(a) 0 0]**.
  If optional parameters <cx> and <cy> are supplied, the rotate is about the point (<cx>, <cy>). The operation represents the equivalent of the following specification: translate(<cx>, <cy>) rotate(<rotate-angle>) translate(-<cx>, -<cy>).

- skewX(<skew-angle>), which specifies a skew transformation along the

x-axis.

- skewY(<skew-angle>), which specifies a skew transformation along the y-axis.

## 8.1.2 Transformation Matrix

Mathematically, all transformations can be represented as 3x3 transformation matrices of the following form:

$$\begin{bmatrix} a & c & e \\ b & d & f \\ 0 & 0 & 1 \end{bmatrix}$$

**Formula 8-1. Transformation Matrix**

Since only six values are used in the above 3x3 matrix, a transformation matrix is also expressed as a vector: [a b c d e f].
Transformations map coordinates and lengths from a new coordinate system into a previous coordinate system:

$$\begin{bmatrix} x_{prevCoordSys} \\ y_{prevCoordSys} \\ 1 \end{bmatrix} = \begin{bmatrix} a & c & e \\ b & d & f \\ 0 & 0 & 1 \end{bmatrix} \cdot \begin{bmatrix} x_{newCoordSys} \\ y_{newCoordSys} \\ 1 \end{bmatrix}$$

**Formula 8-2. Changing Coordinate System**

Simple transformations are represented in matrix form as follows:
Translation is equivalent to the matrix

$$\begin{bmatrix} 1 & 0 & tx \\ 0 & 1 & ty \\ 0 & 0 & 1 \end{bmatrix}$$

**Formula 8-3. Equivalent Matrix of Translation**

or [1 0 0 1 tx ty], where tx and ty are the distances to translate coordinates in X and Y, respectively.
Scaling is equivalent to the matrix

$$\begin{bmatrix} sx & 0 & 0 \\ 0 & sy & 0 \\ 0 & 0 & 1 \end{bmatrix}$$

**Formula 8-4. Equivalent Matrix of Scaling**

or [sx 0 0 sy 0 0]. One unit in the X and Y directions in the new coordinate system equals sx and sy units in the previous coordinate system, respectively.

Rotation about the origin is equivalent to the matrix

$$\begin{bmatrix} \cos(a) & -\sin(a) & 0 \\ \sin(a) & \cos(a) & 0 \\ 0 & 0 & 1 \end{bmatrix}$$

**Formula 8-5. Equivalent Matrix of Rotation**

or [cos(a) sin(a) -sin(a) cos(a) 0 0], which has the effect of rotating the coordinate system axes by angle a.

A skew transformation along the x-axis is equivalent to the matrix

$$\begin{bmatrix} 1 & \tan(a) & 0 \\ 0 & 1 & 0 \\ 0 & 0 & 1 \end{bmatrix}$$

**Formula 8-6. Equivalent Matrix of skewX**

or [1 0 tan(a) 1 0 0], which has the effect of skewing X coordinates by angle a.

A skew transformation along the y-axis is equivalent to the matrix

$$\begin{bmatrix} 1 & 0 & 0 \\ \tan(a) & 1 & 0 \\ 0 & 0 & 1 \end{bmatrix}$$

**Formula 8-7. Equivalent Matrix of skewY**

or [1 tan(a) 0 1 0 0], which has the effect of skewing Y coordinates by angle a.

## 8.2 Implementation approaches

Just like implementing all the elements and attributes above, we have to find an equivalent element or attribute in VML or HTML to do the simulation. The reason why transformation is so difficult to simulate is that for different elements, different approaches should be applied in order to simulate the transformation effects. After some research work, I found several approaches to do the simulation.

### 8.2.1 Using "vml:skew" element

"vml:skew" element can define a transformation for a VML shape. This element has a "matrix" attribute, the value of this attribute is a string in the form "*sxx*, *sxy*, *syx*, *syy*, *px*, *py*" where *s* is scale, *p* is perspective, and *x* and *y* are *x* and *y* values (For more information about how to use *vml:skew* element, please refer to http://msdn.microsoft.com/en-us/library/bb229470(VS.85).aspx). Because "all transformations can be represented as 3x3 transformation matrices" (see last section) we can use this element to simulate SVG transformation. For example, if we want to simulate the transformation *scale(2,1)* in SVG, theoretically, we just need to specify the value of the matrix in "vml:skew" element to be "2, 0, 0, 1, 0, 0" (For more details about how this value comes from, please see the example below).

Before we go further, I must explain something before hand. The transform attribute is applied to an element before processing any other coordinate or length values supplied for that element. For example, in the element

```
<rect x="10" y="10" width="20" height="20" transform="scale(2)"/>
```
**Example 8-1. transform example**

the x, y, width and height values are processed after the current coordinate system has been scaled uniformly by a factor of 2 by the transform attribute. Attributes x, y, width and height (and any other attributes or properties) are treated as values in the new user coordinate system, not the previous user coordinate system. Thus, the above 'rect' element is functionally equivalent to:

```
<g transform="scale(2)">
  <rect x="10" y="10" width="20" height="20"/>
</g>
```
**Example 8-2. transform example**

That is to say, when doing the transformation, not only the shape is transformed, but also the coordinates. Unfortunately, the transformation mechanisms of SVG and "vml:skew" element are different. In order to make it more obvious, I give the complete examples below to explain how I managed to use "vml:skew" element to simulate the transformation effect in SVG. There is not much information or specification about how "<vml:skew>" element works, so the examples below actually result from tests to discover the mechanism of "vml:skew" element little by little.

● Example of "scale"

The SVG elements for testing are as follow:

```
<svg:rect x="0" y="50" width="70" height="50" rx="5"
          fill="none" stroke="purple" stroke-width="10px"/>
<svg:rect x="0" y="150" width="70" height="50" rx="5" fill="none"
          stroke="purple" stroke-width="10px" transform="scale(3,1)"/>
```
**Example 8-3. Testing elements**

The first rectangle is only used for reference. "*scale(3, 1)*" is applied to the second rectangle.

From the specification in 8.1.2, we know that the corresponding matrix for scale is [sx 0 0 sy 0 0] (where sx is the horizontal scale value, sy is the vertical scale value), so the equivalent matrix for "*scale(3, 1)*" is [3 0 0 1 0 0] and the created "`vml:skew`" element should be `<vml:skew matrix="3, 0, 0, 1, 0, 0">`. The result is as follow:

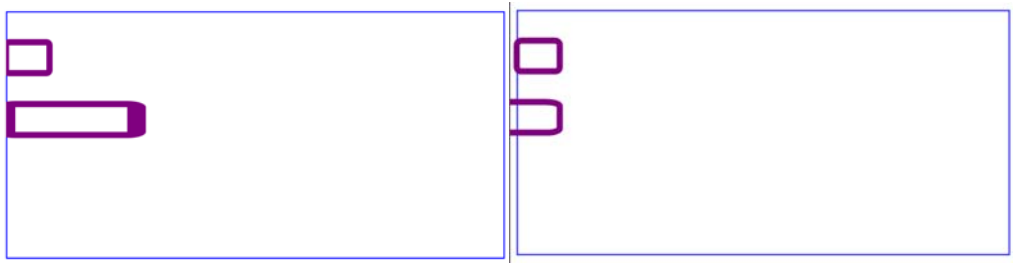

**Figure 8-1. Result in Firefox and IE**

The screen shot on left side is the result in Firefox while the right side one is in IE. From the result we can see that the scaling in SVG starts from the left border of the square and the scaling direction is right-handed. Whereas in IE, the scaling starts from the right border and the scaling direction is left-handed.

In order to convert the direction, I tried the second time by using matrix [-sx 0 0 -sy 0 0] (so the "`vml:skew`" element is now `<vml:skew matrix="-3, 0, 0, -1, 0, 0">`) with the same SVG elements and the result is as follow:

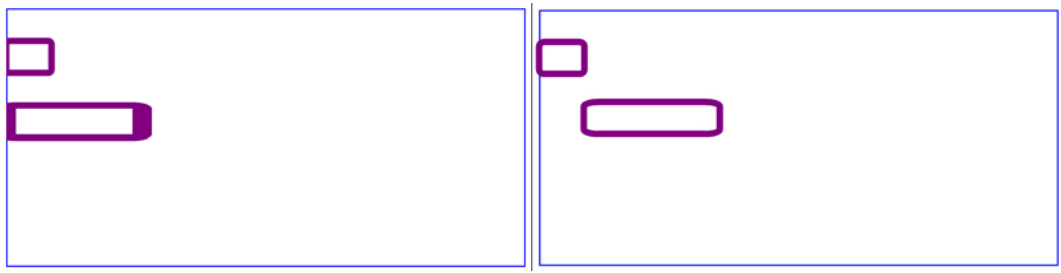

**Figure 8-2. Result in Firefox and IE**

From the result we can see that there is still a difference and the difference is exactly the same to the width of the original rectangle. Therefore, the third test was made by using matrix [-sx 0 0 -sy 0 0], and then move the shape to its left side by the width of the original shape. The result is as follow:

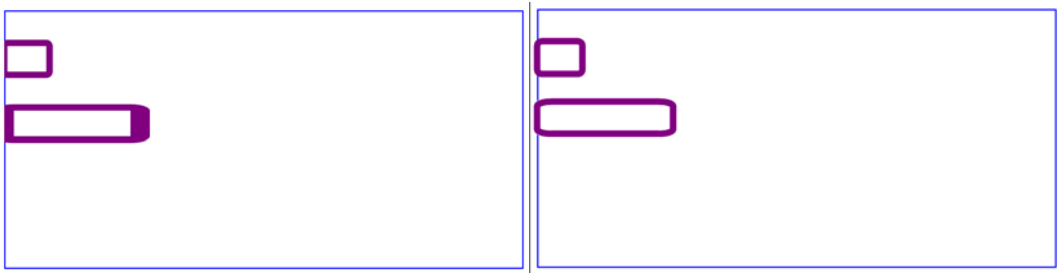

**Figure 8-3. Result in Firefox and IE**

The result seems to be good, but the case is not that simple. When I changed the testing SVG elements from

```
<svg:rect x="0" y="50" width="70" height="50" rx="5"
          fill="none" stroke="purple" stroke-width="10px"/>
<svg:rect x="0" y="150" width="70" height="50" rx="5" fill="none"
          stroke="purple" stroke-width="10px" transform="scale(3,1)"/>
```

**Example 8-4. Original Testing Elements**

into

```
<svg:rect x="100" y="50" width="70" height="50" rx="5" fill="none"
stroke="purple" stroke-width="10px"/>
<svg:rect x="100" y="150" width="70" height="50" rx="5" fill="none"
stroke="purple" stroke-width="10px" transform="scale(3,1)"/>
```

**Example 8-5. New Testing Elements**

That is to move the x coordinate from 0 to 100. The result is as follow:

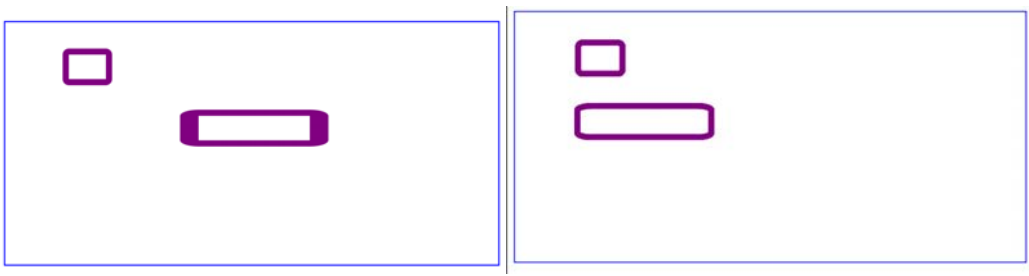

**Figure 8-4. Result in Firefox and IE**

We can see the result is different again. If you are clever enough, you may already have found the reason for this. As I said above "when doing the transformation, not only the shape is transformed, but also the coordinates", if "*scale(3,1)*" is applied to the rectangle, the x coordinates of all the points on that rectangle will also be scaled by 3. This kind of operation is not automatically finished in "vml:skew" so we need to do it manually.

Instead of using the width of the original rectangle to be the offset, I use 3 times of this value to do the last test, and the result is as follow:

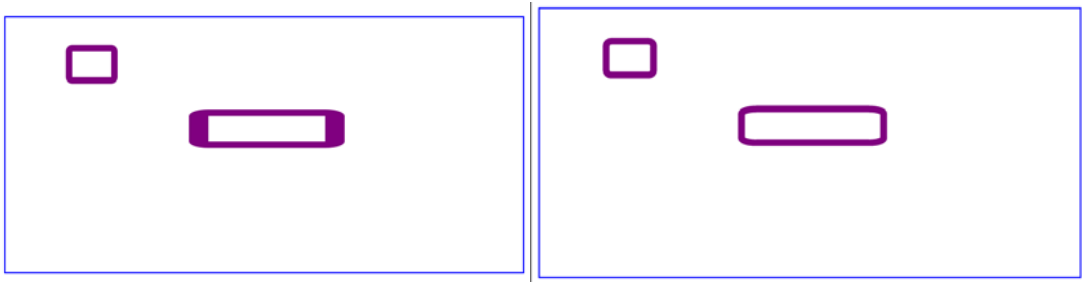

**Figure 8-5. Result in Firefox and IE**

Now the result is perfect. This example only focuses on scaling along X-axis. The implementation of scaling along Y-axis is similar to this example, which is omitted to write.

- Example of "skewX" and "skewY"

The testing SVG elements are as follow:

```
<svg:rect x="0" y="0" width="100" height="100" rx="5" fill="none"
stroke="purple" stroke-width="1px"/>
<svg:rect x="0" y="0" width="100" height="100" rx="5" fill="none"
stroke="blue" stroke-width="1px" transform="skewX(10)"/>
<svg:rect x="0" y="0" width="100" height="100" rx="5" fill="none"
stroke="red" stroke-width="1px" transform="skewX(20)"/>
<svg:rect x="0" y="0" width="100" height="100" rx="5" fill="none"
stroke="yellow" stroke-width="1px" transform="skewX(30)"/>
<svg:rect x="0" y="0" width="100" height="100" rx="5" fill="none"
stroke="black" stroke-width="1px" transform="skewX(40)"/>
```

**Example 8-6. Testing Elements**

This example is similar to the last one. The first rectangle is for reference, the other four are skewed along X-axis by 10, 20, 30, 40 degrees respectively.

From the specification in 8.1.2, we know that the corresponding matrix for skewX is [1 0 tan(a) 1 0 0], so the first test was done based on this matrix and the "vml:skew" elements are:

```
<vml:skew matrix="1, 0, tan(10), 1, 0, 0">
<vml:skew matrix="1, 0, tan(20), 1, 0, 0">
<vml:skew matrix="1, 0, tan(30), 1, 0, 0">
<vml:skew matrix="1, 0, tan(40), 1, 0, 0">
```

**Example 8-7. vml:skew Elements**

The result is as follow:

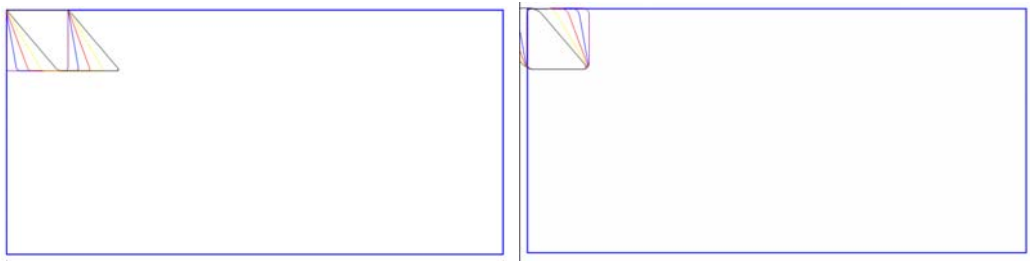

**Figure 8-6. Result in Firefox and IE**

The screen shot on left side is the result in Firefox while the right side one is in IE. From the figure above we can see that the result is also converted: the skewX in SVG starts from the top border of the square and the skewing direction is right-handed. Whereas in IE, the scaling starts from the bottom border and the skewing direction is left-handed.

In order to convert the direction, the second test was made by using matrix [-1 0 -tan(a) -1 0 0] and the result is as follow:

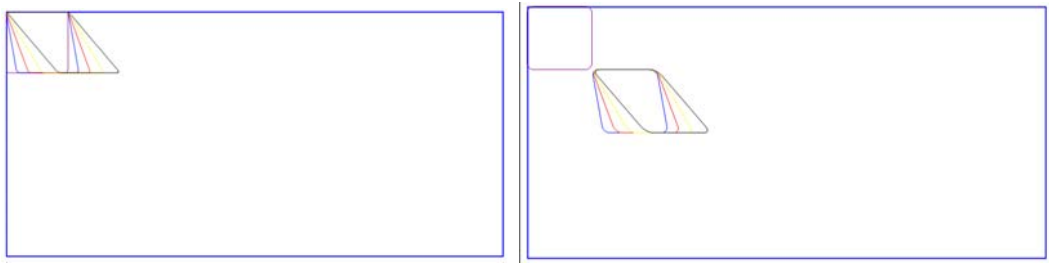

**Figure 8-7. Result in Firefox and IE**

Now the direction is correct, but the offset still exists. From the figure above we can see that the offset values should be as follow:

$$\begin{pmatrix} offsetX \\ offsetY \end{pmatrix} = \begin{pmatrix} width \\ height \end{pmatrix}$$

**Formula 8-8. offsets**

, where "offsetX", "offsetY" are the offset values along x-axis and y-axis respectively. "width" is the width of the original rectangle and "height" is the height of the original rectangle. So we need to move the skewed shapes to the upper left by (offsetX, offsetY), and the result is like this:

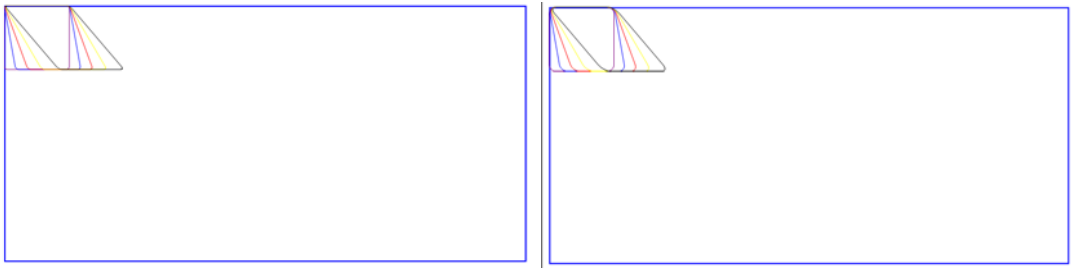

**Figure 8-8. Result in Firefox and IE**

Now the result is fine. But when I changed the testing SVG elements from

```
<svg:rect x="0" y="0" width="100" height="100" rx="5" fill="none"
stroke="purple" stroke-width="1px"/>
<svg:rect x="0" y="0" width="100" height="100" rx="5" fill="none"
stroke="blue" stroke-width="1px" transform="skewX(10)"/>
<svg:rect x="0" y="0" width="100" height="100" rx="5" fill="none"
stroke="red" stroke-width="1px" transform="skewX(20)"/>
<svg:rect x="0" y="0" width="100" height="100" rx="5" fill="none"
stroke="yellow" stroke-width="1px" transform="skewX(30)"/>
<svg:rect x="0" y="0" width="100" height="100" rx="5" fill="none"
stroke="black" stroke-width="1px" transform="skewX(40)"/>
```

**Example 8-8. Original Testing Elements**

into

```
<svg:rect x="0" y="10" width="100" height="100" rx="5" fill="none"
stroke="purple" stroke-width="1px"/>
<svg:rect x="0" y="20" width="100" height="100" rx="5" fill="none"
stroke="blue" stroke-width="1px" transform="skewX(10)"/>
<svg:rect x="0" y="30" width="100" height="100" rx="5" fill="none"
stroke="red" stroke-width="1px" transform="skewX(20)"/>
<svg:rect x="0" y="40" width="100" height="100" rx="5" fill="none"
stroke="yellow" stroke-width="1px" transform="skewX(30)"/>
<svg:rect x="0" y="50" width="100" height="100" rx="5" fill="none"
stroke="black" stroke-width="1px" transform="skewX(40)"/>
```

**Example 8-9. New Testing Elements**

that is to set the "y" attribute to different values, the results are different again:

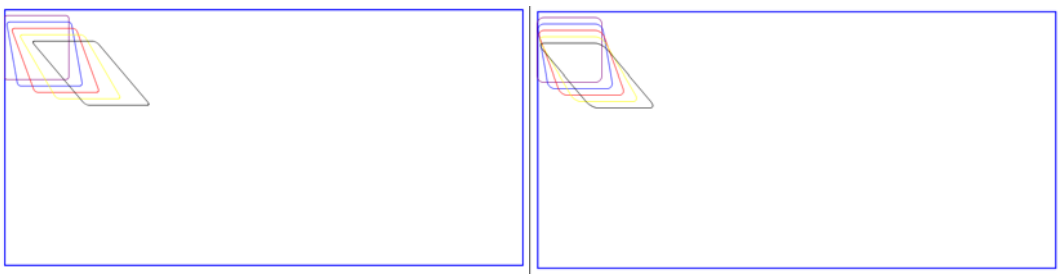

**Figure 8-9. Result in Firefox and IE**

Just like the first example, the changing of the coordinates should also be considered. Therefore, the real offset values should be:

$$\begin{bmatrix} offsetX \\ offsetY \end{bmatrix} = \begin{bmatrix} \tan(a) \cdot top + width \\ height \end{bmatrix}$$

**Formula 8-9. new offsets**

where "top" is the y coordinate of the original shape.

Please note that skewing along X-axis will only affect the X coordinate, so only the offset value along the X-axis is changed.

After moving the shape by the distance of the new offset value, the result is as follow:

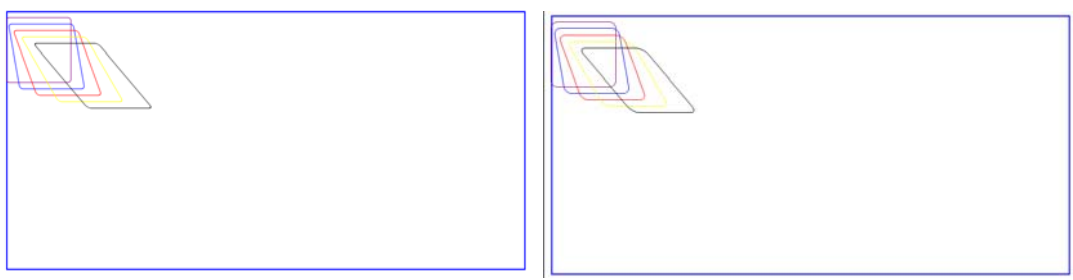

**Figure 8-10. Result in Firefox and IE**

Now the result is identical.

This example only focuses on "skewX". "skewY" is very similar to this example, the transformation matrix for "skewY" is [-1 -tan(a) 0 -1 0 0] and the offset values are:

$$\begin{bmatrix} offsetX \\ offsetY \end{bmatrix} = \begin{bmatrix} width \\ \tan(a) \cdot left + height \end{bmatrix}$$

**Formula 8-10. offsets**

where "left" is the x coordinate of the original shape.

The details of implementing "skewY" are omitted.

- Example of "rotation" and "translate"

The testing SVG elements are:

```
<svg:rect x="350" y="70" width="100" height="200" rx="5" fill="none"
stroke="purple" stroke-width="1px"/>
<svg:rect x="350" y="70" width="100" height="200" rx="5" fill="none"
stroke="grey" stroke-width="1px" transform="rotate(10)"/>
<svg:rect x="350" y="70" width="100" height="200" rx="5" fill="none"
stroke="orange" stroke-width="1px" transform="rotate(20)"/>
<svg:rect x="350" y="70" width="100" height="200" rx="5" fill="none"
stroke="blue" stroke-width="1px" transform="rotate(30)"/>
<svg:rect x="350" y="70" width="100" height="200" rx="5" fill="none"
stroke="red" stroke-width="1px" transform="rotate(40)"/>
<svg:rect x="350" y="70" width="100" height="200" rx="5" fill="none"
stroke="yellow" stroke-width="1px" transform="rotate(50)"/>
<svg:rect x="350" y="70" width="100" height="200" rx="5" fill="none"
stroke="black" stroke-width="1px" transform="rotate(60)"/>
```

**Example 8-10. Testing Elements**

The first rectangle is also for reference, the others are rotated by 10 to 60 degrees respectively.

According to the specification in 8.1.2, the transformation matrix for rotation is [cos(a) sin(a) -sin(a) cos(a) 0 0]. However, the experience of last two examples tells me that, the reference point of rotation in VML is reverted comparing with SVG. Therefore, the matrix should be [-cos(a) -sin(a) sin(a) -cos(a) 0 0] instead. And the result is as follow:

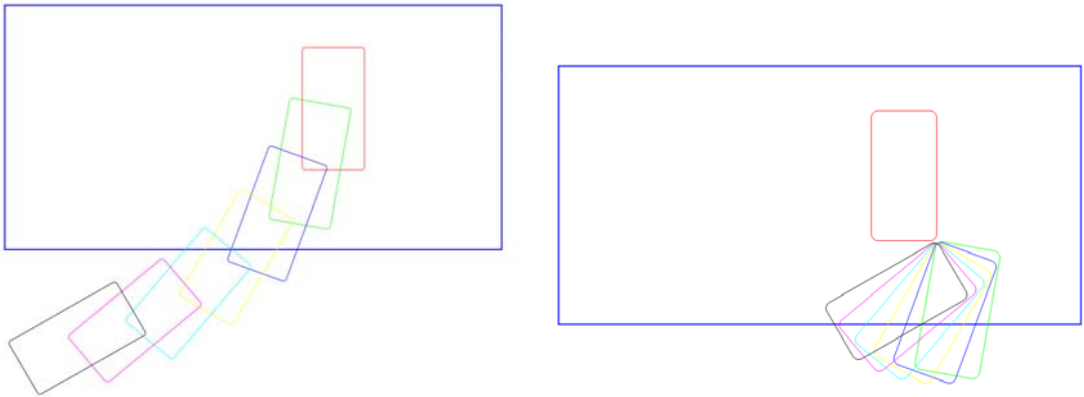

**Figure 8-11. Result in Firefox and IE**

The screen shot on left side is the result in Firefox while the right side one is in IE. We can see the directions and degrees for each rotation are correct but not their offsets.

$$
\begin{bmatrix} offsetX + width \\ offsetY + height \end{bmatrix} = \begin{bmatrix} \cos\theta & -\sin\theta \\ \sin\theta & \cos\theta \end{bmatrix} \begin{bmatrix} x \\ y \end{bmatrix}
$$

**Formula 8-11. offsets**

The formula above is the calculation of the offset values which is a bit more complex than the examples above as both the offsetX and offsetY are changed, and the result is:

$$\begin{bmatrix} offsetX \\ offsetY \end{bmatrix} = \begin{bmatrix} x\cos\theta - y\sin\theta - width \\ x\sin\theta + y\cos\theta - height \end{bmatrix}$$

**Formula 8-12. offsets**

When the offset values are applied, the new result is like follow:

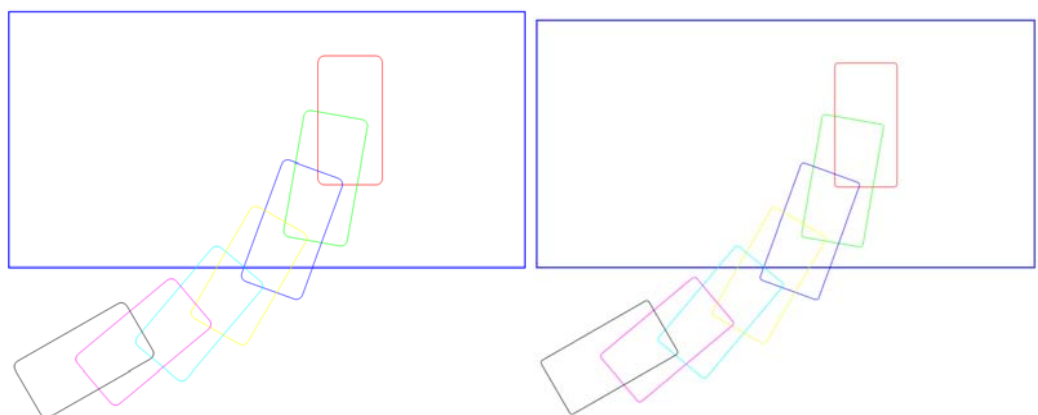

**Figure 8-12. Result in Firefox and IE**

The result is identical now.

In this example, optional parameters <cx> and <cy> are not supplied so the rotation is about the origin of the current user coordinate system. If optional parameters <cx> and <cy> are supplied, the rotation is about the point (<cx>, <cy>). The operation represents the equivalent of the following operation: translate(<cx>, <cy>) rotate(<rotate-angle>) translate(-<cx>, -<cy>).

You may have noticed that I have discussed all the other transformations except "translate". Theoretically, it should be able to use matrix [1 0 0 1 tx ty] to do the transformation. However, the units of the last two parameters in "matrix" attribute of "vml:skew" element are not in pixels, so "translate" is the only transformation that I cannot simulate by using "vml:skew" element. Alternatively, modifying the coordinates is an easier way to do the "translate". For example, "translate(a, b)" can be easily interpreted into:

$$\begin{bmatrix} x' \\ y' \end{bmatrix} = \begin{bmatrix} x + a \\ y + b \end{bmatrix}$$

**Formula 8-13. Translate result**

Then "translate(<cx>, <cy>) rotate(<rotate-angle>) translate(-<cx>, -<cy>)" can be interpreted into:

First, change the current coordinates by (-cx, -cy):

$$\begin{bmatrix} x' \\ y' \end{bmatrix} = \begin{bmatrix} x - cx \\ y - cy \end{bmatrix}$$

**Formula 8-14. Translate result**

Then apply the matrix [-cos(a) -sin(a) sin(a) -cos(a) 0 0] to the shape and calculate the offset values by:

$$\begin{bmatrix} offsetX \\ offsetY \end{bmatrix} = \begin{bmatrix} x\cos\theta - y\sin\theta - width \\ x\sin\theta + y\cos\theta - height \end{bmatrix}$$

**Formula 8-15. offsets**

At last change the coordinates again by (cx, cy):

$$\begin{bmatrix} x' \\ y' \end{bmatrix} = \begin{bmatrix} x + cx \\ y + cy \end{bmatrix}$$

**Formula 8-16. Translate result**

The final result should be:

$$\begin{bmatrix} offsetX \\ offsetY \end{bmatrix} = \begin{bmatrix} \cos\theta(x - cx) - \sin\theta(y - cy) - width + cx \\ \sin\theta(x - cx) + \cos\theta(y - cy) - height + cy \end{bmatrix}$$

**Formula 8-17. offsets**

So far, I have introduced how I use "vml:skew" element to implement the transformation. Briefly speaking, the implementation can be divided into two parts:

1. Applying the "vml:skew" element, which needs to calculate the value of the transform matrix.
2. Calculating the offset values, which needs to know the "x", "y", "width" and "height" values of the shape.

Therefore, "vml:skew" element can only be used to implement the transformation of "circle", "ellipse" and "rect" elements, as only these three elements have "x", "y", "width" and "height" information.

## 8.2.2 Using "CSS Matrix filter"

Some SVG elements such as "foreignObject" and "text" can have child nodes. The child nodes can be either html elements or simply text nodes. We cannot use "vml:skew" element to implement the transformation of these elements because "vml:skew" element must be defined within a VML Shape element whereas the equivalent VML element of "svg:foreignObject" and "svg:text" is "vml:textbox" which is not a VML Shape element.

After some research work, I found "CSS Matrix filter" (I will not explain too much about what "CSS filter" is, for more information about "CSS filter", you can visit http://msdn.microsoft.com/en-us/library/ms532847(VS.85).aspx for more details) is an ideal approach to do the implementation.

The "Matrix Filter" can resize, rotate, or reverse the content of the object using matrix transformation. The following table lists the important attributes:

| Parameters | Description |
|---|---|
| M11 | Sets or retrieves the first row/first column matrix entry for linear transformations. |
| M12 | Sets or retrieves the first row/second column matrix entry for linear transformations. |
| M21 | Sets or retrieves the second row/first column matrix entry for linear transformations. |
| M22 | Sets or retrieves the second row/second column matrix entry for linear transformations. |
| SizingMethod | Sets or retrieves a value that indicates whether the container is resized to fit the resulting image. |

**\*** This table is part from MSDN

**Table 8-1. Matrix Filter Parameters**

From the description above, we can see attributes "M11", "M12", "M21", "M22" are the first four values of the transformation matrix. We can use the same logic in last section to calculate the values of these four parameters and also the offset values if necessary.

Rotation is more complex than the other transformations, so I will only give an example on how I use "Matrix filter" to implement rotation.

The example code is as follow:

```
<svg:foreignObject x="400px" y="300px" width="600px" height="400px">
    <div style="border:1px solid blue;">
        </img>
            <input type="checkbox"></input>
            <input type="radio"></input>
            <textarea style="height:200px; width:300px; font-family:arial;
                             border:1px solid red;">
                Hello this is text
            </textarea>
    </div>
</svg:foreignObject>

<svg:foreignObject x="400px" y="300px" width="600px" height="400px"
transform = "rotate(20, 400, 300)">
    <div style="border:1px solid blue;">
        </img>
            <input type="checkbox"></input>
            <input type="radio"></input>
            <textarea style="height:200px; width:300px; font-family:arial;
                             border:1px solid red;">
                Hello this is text
            </textarea>
    </div>
</svg:foreignObject>
```

**Example 8-11. Testing Elements**

The first "foreignObject" is for reference, and I rotate the second
"foreignObject" by 20 degrees about the point (400, 300).

From the specification, we can calculate the matrix parameters as follow:

$$\begin{bmatrix} M11 \\ M12 \\ M21 \\ M22 \end{bmatrix} = \begin{bmatrix} \cos(20) \\ -\sin(20) \\ \sin(20) \\ \cos(20) \end{bmatrix} = \begin{bmatrix} 0.94 \\ -0.34 \\ 0.34 \\ 0.94 \end{bmatrix}$$

**Formula 8-18. matrix parameter values**

and the "matrix filter" should be as follow:

```
progid:DXImageTransform.Microsoft.Matrix(
M11= 0.94
M12= -0.34
M21= 0.34
M22= 0.94
SizingMethod='auto expand')
```

After applying the matrix filter to the "vml:textbox" element, the result is as follow:

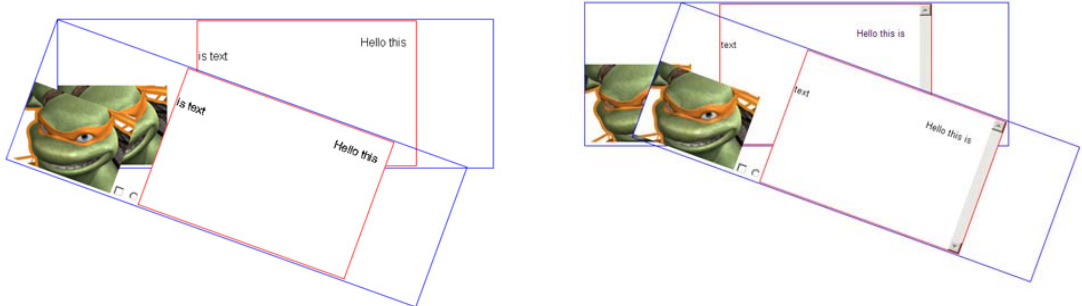

**Screen Shot in Firefox**        **Screen Shot in IE**

**Figure 8-13. Result in Firefox and IE**

The result above tell us "CSS Matrix Filter" is similar to "vml:skew" element which will not calculate the offset values automatically, so we have to calculate the offset values manually. The formula is the same to "vml:skew":

$$\begin{bmatrix} offsetX \\ offsetY \end{bmatrix} = \begin{bmatrix} \cos\theta(x-cx) - \sin\theta(y-cy) - width + cx \\ \sin\theta(x-cx) + \cos\theta(y-cy) - height + cy \end{bmatrix}$$

**Formula 8-19. offsets**

After applying the offset, the final results are the same as follow:

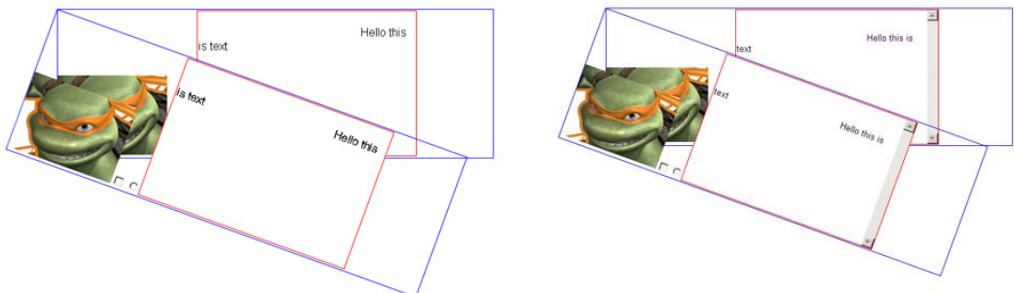

**Screen Shot in Firefox**        **Screen Shot in IE**

**Figure 8-14. Result in Firefox and IE**

Other transformations are easier than the example above so I just omit to write the implementation.

You may ask "is that possible to implement all the other elements by using matrix filter", unfortunately, the answer is no. The following example shows the reason:

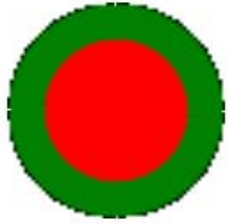

**Figure 8-15. CSS filter on VML shape elements**

The image above is the result of scaling a circle by using "Matrix filter". The side-effect of "Matrix filter" when used on VML shape elements is that it will render raster images, therefore, "Matrix filter" is only applicable to the implementation of "foreignObject", "text" and "textPath" elements.

### 8.2.3 Recalculating the coordinates of the points

The essence of transforming is relocating every point on a specific shape. For the SVG elements which are not defined by points, we must resort to "vml:skew" element or "css matrix filter" to do the implementation which is a bit complex.

However, for elements such as "line", "polyline" and "polygon", we don't have to bother about it. We can simply recalculate the coordinates of every point that helps define a shape.

Like last section, I will only give an example on how to rotate a shape which is defined by "polyline".

The SVG elements for the example are:

```
<svg:polyline fill="none" stroke="blue" stroke-width="5" points="300,300
350,300 350,250 400,250 400,300 450,300" />
<svg:polyline fill="none" stroke="red" stroke-width="5" points="300,300
350,300 350,250 400,250 400,300 450,300" transform="rotate(90, 300, 300)"/>
```

**Example 8-12. Testing Elements**

The first polyline is for reference, the second one is rotated by 90 degrees about point (300, 300).

According to the specification, the formula of rotation about a point (cx, cy) should be:

$$\begin{bmatrix} offsetX \\ offsetY \end{bmatrix} = \begin{bmatrix} \cos\theta(x-cx) - \sin\theta(y-cy) + cx \\ \sin\theta(x-cx) + \cos\theta(y-cy) + cy \end{bmatrix}$$

**Formula 8-20. offsets**

That is to say every point (x, y) that defines the polyline should be calculated by:

$$\begin{bmatrix} offsetX \\ offsetY \end{bmatrix} = \begin{bmatrix} \cos 90(x-300) - \sin 90(y-300) + 300 \\ \sin 90(x-300) + \cos 90(y-300) + 300 \end{bmatrix}$$

**Formula 8-21. offsets**

Then the coordinates of points "300,300 350,300 350,250 400,250 400,300 450,300" will be changed into "300,300 300,350 350,350 350,400 300,400, 300,450". In other words, polyline

```
<svg:polyline fill="none" stroke="red" stroke-width="5" points="300,300
350,300 350,250 400,250 400,300 450,300" transform="rotate(90, 300, 300)"/>
```

**Example 8-13. Polyline Element**

is actually replaced by polyline

```
<svg:polyline fill="none" stroke="red" stroke-width="5" points="300,300
300,350 350,350 350,400 300,400, 300,450"/>
```

**Example 8-14. Polyline Element**

and the result is identical as follow:

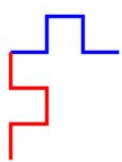
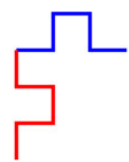

**Screen Shot in IE**    **Screen shot in Firefox**

**Figure 8-16. Result in Firefox and IE**

The implementation of the other transformations (skewX, skewY, translate, scale) are similar to the example above, which I just omit to write.

### 8.2.4 Transformation for <path> element

The three approaches that were introduced above covers most of the SVG elements, however, they are not applicable to the "path" element. The reasons are as follow:

1. It's difficult to retrieve the "x", "y", and "width", "height" values from the

"path" element.

2. Although the "path" element is also defined by a series of points, some commands are mixed in. From the implementation in section 6.3.3 we can see that, it would be too expensive to recalculate all the points.

3. The equivalent VML element to "svg:path" element is "vml:shape" element, so we cannot apply "CSS Matrix Filter" to it otherwise raster images will be rendered.

After some research work, I discovered that it is still possible to use "vml:skew" element to implement the transformation for "path" element although the details of the implementation is a bit different. As I have discussed a lot about "vml:skew" element in section 8.2.1, for more details about how to use "vml:skew" element, you can refer to section 8.2.1. Here I only point out the difference between the two implementations:

1. Unlike in section 8.2.1, the transformation matrix for "vml:shape" element needn't to be reverted.

2. The offset values are not determined by "X", "Y" and "width", "height"values, but by "rootWidth" (width of SVG root element) and "rootHeight" (height of SVG root element) values.

I will give another example to show how I managed to rotate a "path" element. The example code is as follow:

```
<svg:path d="M 100 100 200 200 L 300 100  200 300 Z " fill="red" stroke="blue"
stroke-width="3"/>
<svg:path d="M 100 100 200 200 L 300 100  200 300 Z " fill="red" stroke="blue"
stroke-width="3" transform="rotate(20, 300, 300)"/>
```

**Example 8-15. Testing Elements**

The first path is for reference, the second one is rotated by 20 degree about point (300, 300).

The steps of implementing the transformation are:

1. Applying the transformation matrix [cos(angle), -sin(angle), sin(angle), cos(angle), 0, 0] to "vml:skew" element.

2. Calculating the offset values by:

$$\begin{bmatrix} offsetX \\ offsetY \end{bmatrix} = \begin{bmatrix} \cos\theta \cdot rootWidth - \sin\theta \cdot rootHeight - rootWidth \\ \sin\theta \cdot rootWidth + \cos\theta \cdot rootHeight - rootHeight \end{bmatrix}$$

**Formula 8-22. offsets**

The other transformations are easier than the example above so I just omit to mention them.

### 8.2.5 Transformation for <g> element

When a transformation is applied to a "g" element, all the elements nested in the "g" elements will inherit this transformation. For example, the result of

```
<g transform="scale(2)">
  <rect x="10" y="10" width="20" height="20"/>
  <line x1="0" y1="0" x2="50" y2="0" />
</g>
```
**Example 8-16. Testing Elements**

is actually equal with

```
<g>
  <rect x="10" y="10" width="20" height="20" transform="scale(2)"/>
  <line x1="0" y1="0" x2="50" y2="0" transform="scale(2)"/>
</g>
```
**Example 8-17. Testing Elements**

Therefore, no extra elements or attributes are required to implement the transformation of "g" element, what we only need to do is to apply the transformation of "g" element to all its sub elements and then have the sub elements be rendered by one of the approaches mentioned above.

## 8.3 Multiple-transformations

The value of the transform attribute can be either a single transformation or a transform-list which is defined as a list of transform definitions. The individual transform definitions are separated by white space and/or a comma.

Transform-list can consist of any number of transformations. The effect of transform-list is to post-multiply the subsequent transformation matrices onto previously defined transformations. For each given element, the accumulation of all transformations that have been defined on the given element establishes the current transformation matrix or CTM. The CTM thus is as follow:

$$CTM = \begin{bmatrix} a_1 & c_1 & e_1 \\ b_1 & d_1 & f_1 \\ 0 & 0 & 1 \end{bmatrix} \cdot \begin{bmatrix} a_2 & c_2 & e_2 \\ b_2 & d_2 & f_2 \\ 0 & 0 & 1 \end{bmatrix} \cdot \ldots \cdot \begin{bmatrix} a_n & c_n & e_n \\ b_n & d_n & f_n \\ 0 & 0 & 1 \end{bmatrix}$$

**Formula 8-23. CTM**

Therefore, when implementing the multi-transformation, I considered the transform-list as a single transform. I first calculate the CTM from the transform-list and then all the rest of the work will be the same as single-transformation.

Unfortunately, the offset calculation of multi-transformation is more complex than single transformation so I didn't succeed in finding the solutions for all the multi-transformations when I was writing this essay.

## 8.4 Summary of how transformations have been implemented

Implementing transformation was interesting but difficult. I managed to implement all the single transformations but the multi-transformation is not finished yet. The table below shows the current status of the implementation:

| Approaches | | Using <vml:skew/> | Using <vml:skew/> | Recalculating points | CSS Matrix filter | Apply to child nodes |
|---|---|---|---|---|---|---|
| Applicable elements | | <circle/> <ellipse/> <rect/> | <path/> | <line/> <polyline/> <polygon/> | <foreignObject/>  <textPath/> | <g/> |
| Transformation Implementation Status | Single | Yes | Yes | Yes | Yes | Yes |
| | Multiple | Implemented "scale", "translate", "skewX", "skewY" | Implemented "scale", "translate" | All | Implemented "scale", "skewX", "skewY" | Depends on child node types |

**Table 8-2. Transformation Implementing Status**

# 9. Conclusion

SVG is a W3C specification that has a lot of powerful features. In this project, I created a Javascript file (according to the format of Backbase Client Framework binding) that translates SVG images to VML images displayed without need for a plug-in in Internet Explorer. It supports SVG document structure elements such as "svg", "g", "def", "use". It also supports all SVG geometric shape elements as well as text elements such as "rect", "circle", "ellipse", "line", "polyline", "polygon" and "text", "textPath". Besides, the SVG "path" element is also implemented and can be used to draw complex shapes such as Bezier Curves. Moreover, some very nice feathers such as gradient and transformation are also implemented in this project.

However, this project only implemented a sub-set of SVG elements compared to the full capability of SVG. Some further work still needs to be done:

**Multi-transformation**
The multi-transformation is not fully implemented yet because of the complexity of the calculation of the offset values. Just like when I implemented the single transformation, there is no specification about how VML or CSS Marix Filter works, so all my current solutions come from plenty of test work. I believe there must be a solution for all the other multi-transformations and that is only a matter of time before it will be solved.

**"a" command of <path> element**
"a" command of SVG "path" element is a very important command which can be used to generate 3D pie charts. The most suitable VML command for mapping this command seems to be "at" command. However, the implementation still reminds to be a very difficult question due to the difference between the parameters of these two commands.

**Other aspects**
This project doesn't support visual effects like blending images, dropping shadows, diagonal gradients, and multi-level grouped elements. Gradients in SVG can contain more than two colors; in the meantime, radialGradient is still missing. These are very nice features but I haven't found a solution to implement any of them.

# 10. Acknowledgements

During the whole project, many people gave me a lot of good advices.

Appreciation to Dr. Nies Huijsmans, who is my project advisor in Leiden University. He gave me many good advices and helped me to correct this paper.

Appreciation to Mr. Dan Tsymbala and Mr. Sergey Ilinsky, who are my advisors in Backbase. They let me know how powerful the framework is and gave me a lot of useful instructions on programming.

Appreciation to Ms. Dimitra, who is responsible for my internship in Backbase. She helped me to organize my presentation and correct my essay.

Appreciation to Backbase, which gave me the opportunity to finish my master's project.

# Bibliography

- **W3C - Scalable Vector Graphics (SVG) 1.1 Specification**

  http://www.w3.org/TR/SVG11/

- **W3C - Vector Markup Language (VML)**

  http://www.w3.org/TR/NOTE-VML

- **W3C - Synchronized Multimedia Integration Language (SMIL 2.1)**

  http://www.w3.org/TR/SMIL2/

- **W3C - XML Linking Language (XLink) Version 1.0**

  http://www.w3.org/TR/xlink/

- **Mozilla - Mozilla Developer Center Core JavaScript 1.5 Reference**

  https://developer.mozilla.org/en/Core_JavaScript_1.5_Reference

- **Mozilla - SVG in Firefox-Element implementation status**

  http://developer.mozilla.org/en/SVG_in_Firefox

- **Opera - SVG support in Opera**

  http://www.opera.com/docs/specs/opera95/svg/

- **Safari - SVG support in Safari**

  http://webkit.org/projects/svg/status.xml

- **MSDN - Vector Markup Language (VML)**

  http://msdn.microsoft.com/en-us/library/bb250524.aspx

- **Google - Google Charts**

  http://code.google.com/apis/chart/

- **Backbase - Application Development Guide**

- **Backbase - Widget Development Guide**

- **Backbase – Backbase Makeup language & Backbase API Reference**

- **Stylesheet Translations of SVG to VML**

  - Julie Nabong San, Jose State University, May 2004

- **Adobe SVG Project**

  http://www.adobe.com/svg/

- **How To Select an Implementation for Vector Graphics Across Browsers**

  - Liang Wang, Leiden University, Nov 2008

# Appendix A – Glossary

- **Backbase and Backbase Client Framework**

  Backbase is the leading provider of enterprise software for creating AJAX-based Rich Internet Applications (RIAs).

  The Backbase Client Framework is a lightweight Client Runtime embedded in the client web browser, the Client Framework provides a complete suite of reusable, extensible widgets and an extensive events system that allow web developers to efficiently build rich AJAX applications.

- **HTML (HyperText Markup Language)**

  It provides an approach to describe the structure of text-based information in a document - by denoting certain text as links, headings, paragraphs, lists and to supplement that text with interactive forms, embedded images, and other objects.

- **JavaScript**

  A scripting language that can be embedded on Web documents to add interactivity.

- **DOM (Document Object Model)**

  It is a platform and language independent standard object model for representing HTML or XML documents as well as an Application Programming Interface (API) for querying, traversing and manipulating such documents.

- **SVG (Scalable Vector Graphics)**

  It is an XML-based application to create two-dimensional images useful for charts, graphs, and statistical data.

- **VML (Vector Markup Language)**

It is an XML-based application to create two-dimensional images which can be viewed in Internet Explorer 5.0 and higher.

- **W3C (World Wide Web Consortium)**

  An organization created in 1994 that develops interoperable software, guidelines, tools, and specifications to make the Web available to all users.

- **XML (Extensible Markup Language)**

  A simple, very flexible text format derived from SGML. Originally designed to meet the challenges of large-scale electronic publishing, XML is also playing an increasingly important role in the exchange of a wide variety of data on the Web and elsewhere.

- **CSS (Cascading Style Sheets)**

  A is a stylesheet language used to describe the presentation (that is, the look and formatting) of a document written in a markup language.

- **SMIL (Synchronized Multimedia Integration Language)**

  It is a W3C recommended XML markup language for describing multimedia presentations. It defines markup for timing, layout, animations, visual transitions, and media embedding, among other things.

- **XEL (XML Execution Language)**

  A XEL is a programming language that application developers will use with the Backbase Client Framework product to define handlers and execute functional logic on instances in their application.

# Appendix B - SVG elements and their VML counterparts

| | FF 2.0 | FF 3.0 | OP 9.6 | SF 3.0 | VML\| HTML equivalent (IE) | Backbase Client Framework |
|---|---|---|---|---|---|---|
| | | | **Implemented Elements** | | | |
| svg (W3C) | partial/incomplete | partial/incomplete | **yes** | partial/incomplete | group (MSDN) | **yes** |
| g (W3C) | **yes** | **yes** | **yes** | partial/incomplete | group (MSDN) | **yes** |
| defs (W3C) | **yes** | **yes** | **yes** | **yes** | html:div | **yes** |
| use (W3C) | partial/incomplete | partial/incomplete | partial/incomplete | **yes** | html:div | **yes** |
| circle (W3C) | **yes** | **yes** | **yes** | **yes** | oval (MSDN) | **yes** |
| ellipse (W3C) | **yes** | **yes** | **yes** | **yes** | oval (MSDN) | **yes** |
| path (W3C) | partial/incomplete | partial/incomplete | **yes** | **yes** | shape (MSDN) | **yes** |
| rect (W3C) | **yes** | **yes** | **yes** | **yes** | roundrect (MSDN) | **yes** |
| line (W3C) | **yes** | **yes** | **yes** | **yes** | shape (MSDN) | **yes** |
| polygon (W3C) | **yes** | **yes** | **yes** | **yes** | shape (MSDN) | **yes** |
| polyline (W3C) | **yes** | **yes** | **yes** | **yes** | shape (MSDN) | **yes** |
| text (W3C) | partial/incomplete | partial/incomplete | **yes** | **yes** | textbox (MSDN) | **yes** |
| textPath (W3C) | partial/incomplete | partial/incomplete | **yes** | **yes** | textPath (MSDN) | **yes** |
| linearGradient (W3C) | **yes** | **yes** | **yes** | partial/incomplete | fill (MSDN) | **yes** |
| stop (W3C) | **yes** | **yes** | **yes** | **yes** | fill (MSDN) | **yes** |
| a (W3C) | partial/incomplete | partial/incomplete | **yes** | **yes** | html:a | **Yes** |
| foreignObject (W3C) | no | yes | partial/incomplete | yes | textbox (MSDN) | Yes |

**Already Implemented by the Framework in other approaches**

| | | | | | |
|---|---|---|---|---|---|
| image (W3C) | partial/incomplete | partial/incomplete | partial/incomplete | partial/incomplete | Not necessary to be implemented |
| style (W3C) | yes | yes | yes | yes | Not necessary to be implemented |
| script (W3C) | yes | yes | partial/incomplete | partial/incomplete | Not necessary to be implemented |
| Font Module | partial/incomplete | partial/incomplete | partial/incomplete | partial/incomplete | Not necessary to be implemented |
| Animation Module | partial/incomplete | partial/incomplete | partial/incomplete | partial/incomplete | Not necessary to be implemented |

**No practical use elements**

| | | | | | |
|---|---|---|---|---|---|
| desc (W3C) | partial/incomplete | partial/incomplete | yes | yes | Not going to implement |
| title (W3C) | yes | yes | yes | yes | Not going to implement |
| metadata (W3C) | partial/incomplete | partial/incomplete | yes | yes | Not going to implement |
| symbol (W3C) | yes | yes | yes | yes | Not going to implement |
| switch (W3C) | yes | yes | yes | yes | Not going to implement |
| radialGradient (W3C) | yes | yes | yes | partial/incomplete | Not going to implement |
| cursor (W3C) | no | no | no | partial/incomplete | Not going to implement |

**Elements/Modules cannot find a solution**

| | | | | | |
|---|---|---|---|---|---|
| tref (W3C) | no | no | yes | yes | No solution |
| tspan (W3C) | partial/incomplete | partial/incomplete | yes | yes | No solution |
| [altGlyph (W3C)](#) | *no* | *no* | *no* | *no* | *No solution* |
| [altGlyphDef (W3C)](#) | *no* | *no* | *no* | *no* | *No solution* |
| [altGlyphItem (W3C)](#) | *no* | *no* | *no* | *no* | *No solution* |
| [glyphRef (W3C)](#) | *no* | *no* | *no* | *no* | *No solution* |
| [marker (W3C)](#) | **yes** | **yes** | **yes** | **yes** | *No solution* |
| [color-profile (W3C)](#) | *no* | *no* | *no* | *no* | *No solution* |
| [pattern (W3C)](#) | *no* | **yes** | **yes** | **yes** | *No solution* |
| [clipPath (W3C)](#) | partial/incomplete | partial/incomplete | **yes** | **yes** | *No solution* |
| [mask (W3C)](#) | *no* | **yes** | **yes** | **yes** | *No solution* |
| [view (W3C)](#) | *no* | *no* | **yes** | partial/incomplete | *No solution* |
| Filter Module | partial/incomplete | partial/incomplete | partial/incomplete | partial/incomplete | *No solution* |